\documentclass[sigconf,nonacm]{acmart}
\AtBeginDocument{%
  }

\setcopyright{none}             
\renewcommand\footnotetextcopyrightpermission[1]{} 

\usepackage{booktabs}  
\usepackage{tabularx}  
\usepackage{caption}   
\usepackage{multirow} 
\usepackage{amsmath}  
\usepackage{xcolor}   
\usepackage{wrapfig}

\begin{document}

\title{A Multi-Asset Dataset and Tail-Risk Framework for DeFi Yield Trading Derivatives}

\author{Huaiyu Jia}
\affiliation{%
  \institution{The Hong Kong University of Science and Technology (Guangzhou)}
  \city{Guangzhou}
  \country{China}
}
\email{hjia351@connect.hkust-gz.edu.cn}

\author{Jieshun You}
\affiliation{%
  \institution{Hong Kong Polytechnic University}
  \city{Hong Kong}
  \country{Hong Kong SAR}
}
\email{24041917g@connect.polyu.hk}

\author{Jingyu Liu}
\affiliation{%
  \institution{The Hong Kong University of Science and Technology (Guangzhou)}
  \city{Guangzhou}
  \country{China}
}
\email{jliu514@connect.hkust-gz.edu.cn}

\author{Yizhi Luo}
\affiliation{%
  \institution{The Hong Kong University of Science and Technology (Guangzhou)}
  \city{Guangzhou}
  \country{China}
}
\email{yluo198@connect.hkust-gz.edu.cn}

\author{Shuo Sun}
\authornote{Corresponding author.}
\affiliation{%
  \institution{The Hong Kong University of Science and Technology (Guangzhou)}
  \city{Guangzhou}
  \country{China}
}
\email{shuosun@hkust-gz.edu.cn}


\begin{abstract}
  Yield trading has emerged as a pivotal innovation in Decentralized Finance (DeFi) by decoupling principal from yield streams to enhance capital efficiency. Despite its significance, the field lacks fine-grained datasets for systematic empirical analysis. We address this gap by constructing the first large-scale benchmark dataset for Pendle—the pioneering on-chain interest rate derivatives protocol—spanning 24 months across 340 active markets with complete lifecycle observations. Our analysis reveals a structural market transition toward stablecoin-dominated "flight-to-quality" dynamics. To tackle the heavy-tailed nature of yield data, we propose Tail-MSE, an asymmetric loss function that prioritizes tail risk awareness through dynamic penalty weighting. Experiments across six state-of-the-art time series models demonstrate consistent improvements in volatility forecasting robustness, enabling the Tail-MSE Sniper strategy to achieve superior risk-adjusted returns in over half of all markets.
\end{abstract}



\maketitle

\section{INTRODUCTION}

Decentralized Finance (DeFi) \cite{auer2023technology} has reshaped traditional finance through permissionless innovations like decentralized exchanges \cite{coingecko_dex} and on-chain lending \cite{defillama_lending}. Among these, yield trading \cite{defillama_yield} has emerged as a pivotal paradigm by dividing assets into principal and yield streams, enabling both passive income and active yield speculation while enhancing capital efficiency.

\begin{figure}[htbp]
    \centering
    \includegraphics[width=0.95\linewidth]{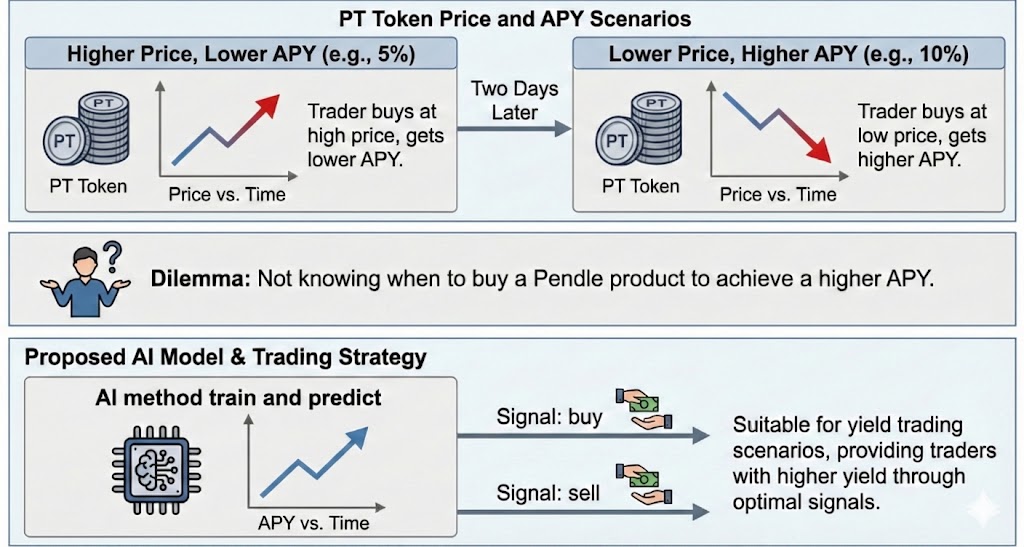}
    \caption{Motivation: Pendle Yield Trading Dilemma}
    \label{fig:Motivation}
\end{figure}

\begin{figure}[htbp]
    \centering
    \includegraphics[width=0.95\linewidth]{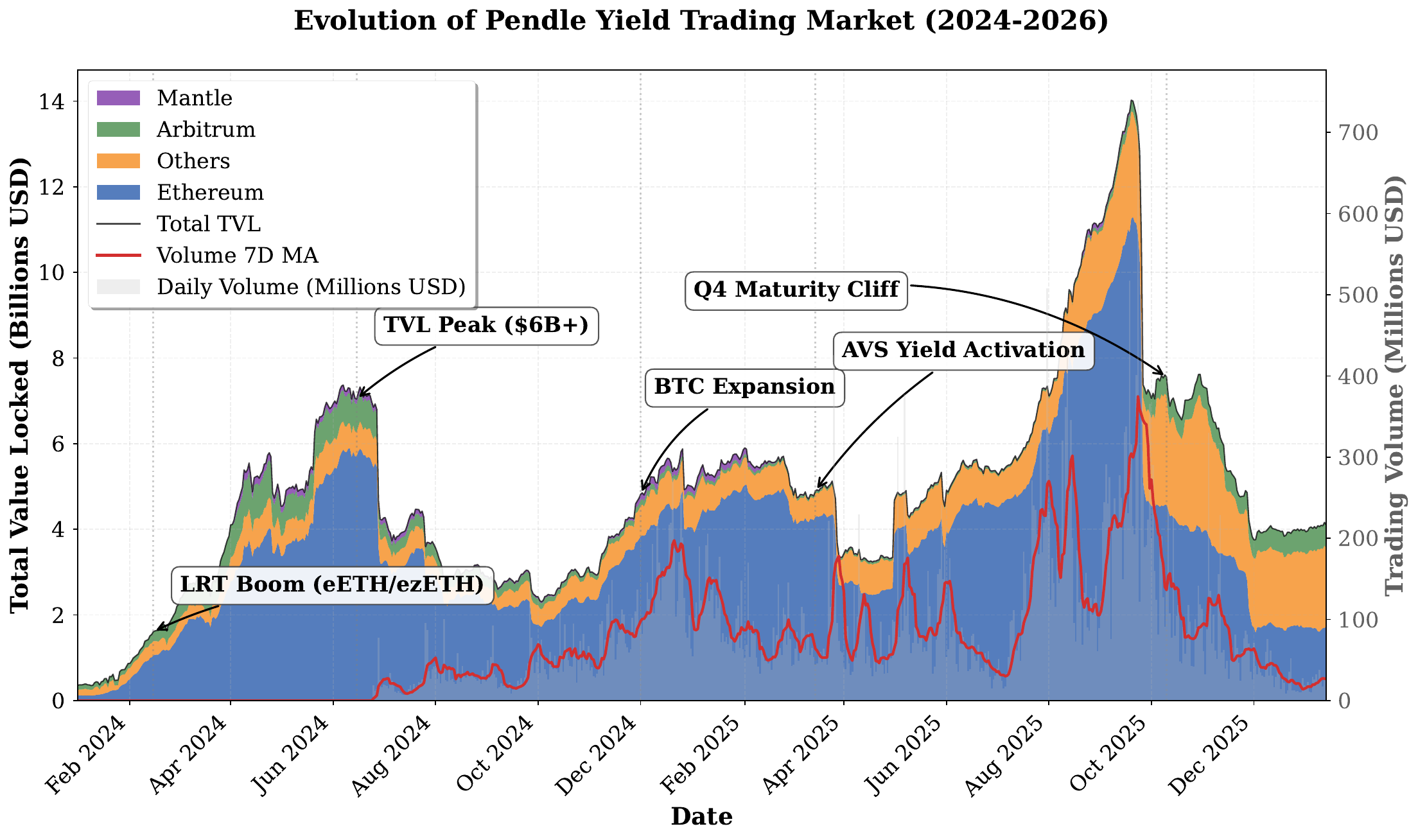}
    \caption{Evolution of Pendle Yield Trading Market Dynamics}
    \label{fig:pendleAnalysis}
\end{figure}

Pendle \cite{pendle_finance} is an on-chain interest rate derivatives protocol that has developed a substantial trading ecosystem, as shown in Figure \ref{fig:pendleAnalysis}, making it a representative platform for studying yield trading. The protocol employs a time-decaying automated market maker (AMM) to split yield-bearing assets into Principal Tokens (PT) \cite{pendle_pt} and Yield Tokens (YT) \cite{pendle_yt}. As illustrated in Figure \ref{fig:Motivation}, traders face a fundamental trade-off: PT price and the implied annual percentage yield (APY) move in opposite directions. For example, buying PT at a high price locks in a lower yield (e.g., 5\%), only to see the APY rise significantly (e.g., to 10\%) as the PT price later declines. This dynamic highlights the need for predictive models that can capture the temporal behavior of yields.

Although the yield trading field has expanded rapidly, previous research remains limited due to a lack of high-quality, event-level datasets that capture yield decomposition mechanics and token time-decay dynamics \cite{xu2022met,fan2021strategic}. Existing datasets primarily focus on spot markets (e.g., Uniswap) \cite{uniswap2021v3} and do not record the complete lifecycle of yield trading—from minting to redemption. This data gap constrains empirical analysis of yield trading market behavior.

To address the data gap in yield trading research, we construct a fine-grained  dataset focused on this domain. Most existing yield trading protocols suffer from low market capitalization and limited liquidity, rendering them unsuitable for systematic study \cite{panews2025notional}; Pendle stands out as the only protocol with sustained activity and substantial trading volume, making it an appropriate subject for empirical analysis. Our dataset spans two years (Jan 2024 to Jan 2026) and covers 340 markets across multiple underlying asset classes, including stablecoins, ETH-Beta, BTC-Fi and son on. Each market is uniquely identified by its maturity date and smart contract addresses, with hourly market snapshots. This resource enables the first comprehensive empirical investigation of yield trading. 

Based on the dataset, we set out to address the challenges of predictive modeling in markets with moderate volatility and non-Gaussian data distribution. We propose a novel Tail-MSE (Tail-sensitive Mean Squared Error) loss function designed to mitigate the risks associated with the heavy-tailed nature of Pendle's product data. Unlike standard MSE, which treats positive and negative errors symmetrically, Tail-MSE incorporates an asymmetric penalty mechanism that prioritizes the learning of extreme downside risks and volatility spikes. This methodological innovation aligns the model's objective with the financial imperative of capital preservation. Furthermore, we translate these predictive capabilities into a robust quantitative trading strategy. By leveraging the model's enhanced sensitivity to tail events, the proposed strategy demonstrates the ability to navigate the inherent long-tail risks of yield products, offering actionable guidance for optimizing risk-adjusted returns in yield trading.

We summary our contributions as below:

\begin{itemize}
    \item First Fine-Grained Pendle Yield Trading Dataset: We construct a benchmark dataset spanning 24 months (January 2024–January 2026) with 896,811 hourly observations across 340 markets, systematically stratified into four asset classes—Stablecoins (44.3\%), ETH-Beta (24.2\%), BTC-Fi (10.8\%), and Others (20.7\%)—filling a critical void in yield derivative empirical research.
    \item Tail-Risk-Aware Prediction Methodology: We propose Tail-MSE, an asymmetric loss function with dynamic penalty weighting $\alpha$ that prioritizes tail risk learning and validated across six time series models.
    \item Quantitative Strategies for Yield Trading: We design the Tail-MSE Sniper strategy that outperforms baselines in 181 markets (53.2\%) with 7.8\% median ROI and only 3.7\% maximum drawdown, achieving a 69.2\% win rate in Stablecoin markets and delivering the first executable algorithmic framework for yield trading assets.
\end{itemize}

\section{RELATED WORK}
\subsection{DeFi Yield Derivatives}
The focus of decentralized finance (DeFi) research has evolved significantly, transitioning from asset-level price analysis \cite{long2025bridging} to complex protocol-level yield dynamics \cite{zhou2025graph}. Early literature modeled price movements across different asset classes, encompassing basic cryptocurrencies \cite{lahmiri2021deep, sebastiao2021forecasting}, perpetual futures \cite{chen2024perpetual}, and the recently emerging Meme tokens \cite{mongardini2025midsummer}. Beyond heuristic-based anomaly detection, recent research has increasingly leveraged on-chain transaction data \cite{luo2024ai} to build deep learning models for transaction cost optimization and alpha generation. Subsequently, the research paradigm shifted to yield-generating protocols. Extensive empirical analyses of Uniswap V3 liquidity supply \cite{heimbach2022risks} and the Aave lending market \cite{aave2026} demonstrate a growing academic interest in quantifying capital efficiency and the stochastic nature of liquidity mining strategies.

However, despite these advances, research on "yield trading"—the derivatives market for speculating on future interest rates—remains extremely scarce. Existing literature on yield derivatives is largely limited to theoretical mechanisms \cite{nadkarni2025split,bastankhah2024thinking}, lacking fine-grained, data-driven empirical analysis. Furthermore, given that many emerging yield protocols have collapsed due to flawed economic models or liquidity crises \cite{ainvest2025notional} , the resilience and sustained market activity exhibited by the Pendle protocol make it particularly unique.

\subsection{Quantitative Strategies in DeFi}
Cryptocurrency price prediction and yield modeling have evolved from traditional machine learning \cite{mujlid2023survey} and deep learning \cite{yi2023frequency} to Large Language Model (LLM)-based agents \cite{li2024cryptotrade}, and the research scope has expanded from single-asset price prediction \cite{wang2023ex} to portfolio optimization \cite{guan2025security}, risk hedging, and high-frequency trading \cite{han2025pulsereddit}. However, existing work rarely focuses on the modeling and trading of on-chain protocol yield \cite{xu2022reap}. A few studies focus on maximizing liquidity mining yield and propose heuristic algorithms \cite{ni2024money}; other works analyze the risk-reward trade-offs of liquidity providers in AMM protocols such as Uniswap \cite{milionis2022automated}. Currently, there is still a significant gap in quantitative strategy research for yield trading protocols such as Pendle, whose dual-token structure (principal/yield token) creates complex nonlinear pricing dynamics. Existing models struggle to capture the unique arbitrage and hedging opportunities inherent in this "fixed-yield" DeFi layer. This paper aims to fill this gap by proposing a customized trading strategy based on fine-grained protocol data.

\begin{figure}[htbp]
    \centering
    \includegraphics[width=0.95\linewidth]{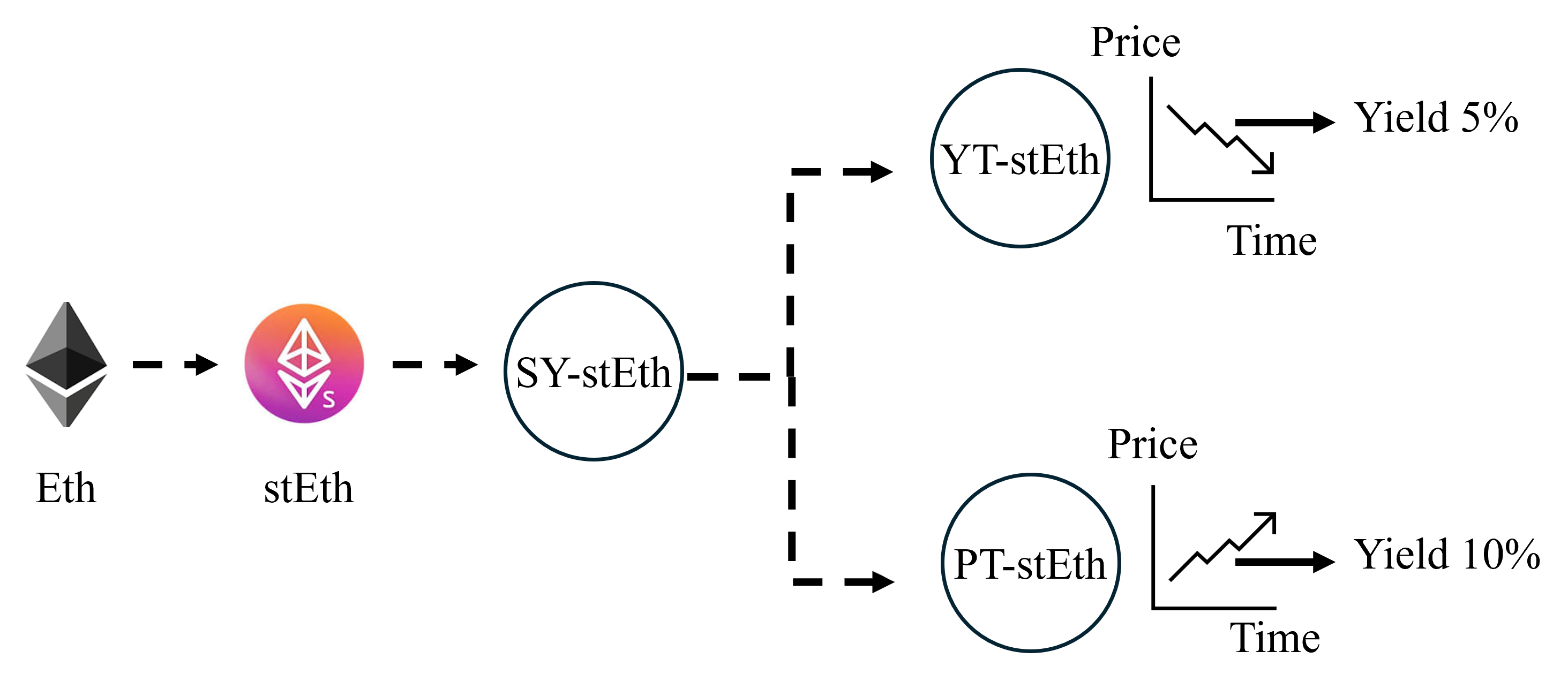}
    \caption{Pendle Yield  Tokenization Mechanism}
    \label{fig:tokenFlow}
\end{figure}

\begin{figure*}
    \centering
    \includegraphics[width=0.95\textwidth]{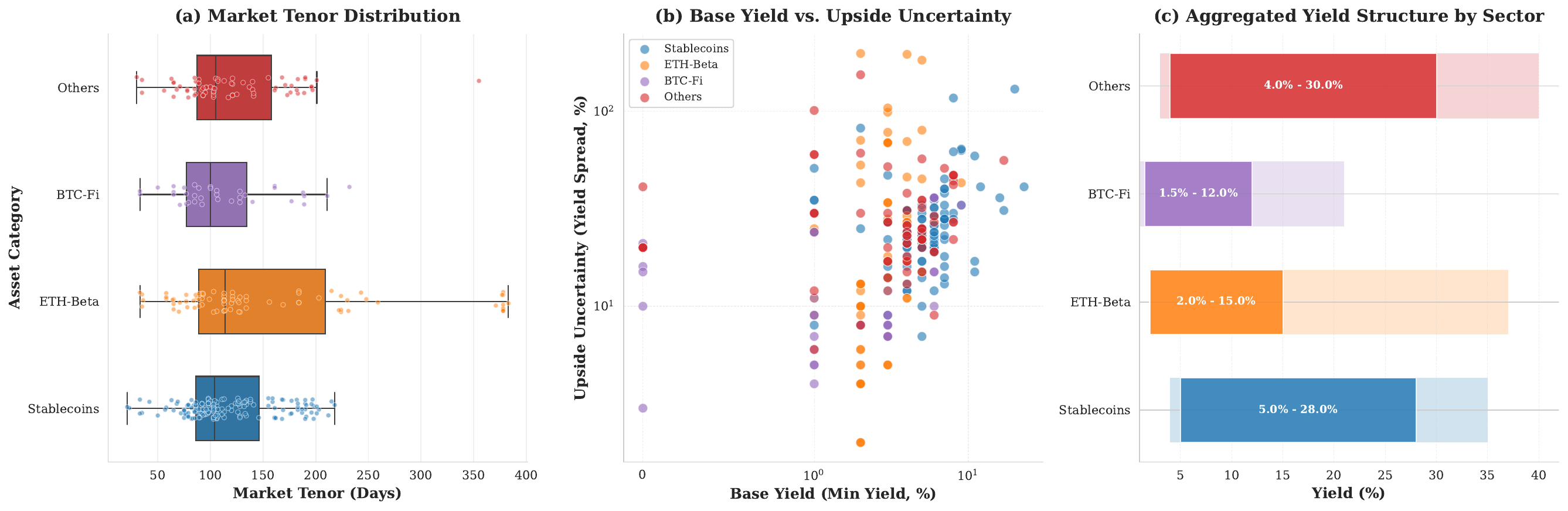}
    \caption{Multi-dimensional Empirical Analysis of Market Tenors and Yield Structures}
    \label{fig:combined}
\end{figure*}

\section{DATASET}
\subsection{Dataset Formulation}
The Pendle protocol operates on a finite-maturity framework with expiry date $T$. Underlying assets (e.g., stETH) are standardized into SY tokens and decomposed into two derivatives satisfying $P_{SY}(t) = P_{PT}(t) + P_{YT}(t)$. As $t \to T$, $P_{PT}$ converges to one while $P_{YT}$ decays to zero, yielding two distinct instruments:

\textbf{Principal Token (PT)} functions as a zero-coupon bond trading below par ($P_{PT}(t) < 1$), locking in fixed returns through guaranteed appreciation to face value at maturity. Lower $P_{PT}$ values correspond to higher potential returns.

\textbf{Yield Token (YT)} provides leveraged exposure to variable on-chain rewards, with profitability determined by the difference between its upfront cost ($P_{YT}(t)$) and the actual yield accumulated before maturity.

\textbf{Implied APY} serves as a forward-looking market signal derived from PT pricing, defined as $\mathrm{APY} = \left( \frac{1}{P_{PT}(t)} - 1 \right)^{1/\tau} - 1$, where $\tau$ denotes time-to-maturity. This metric provides a real-time, data-driven signal of market expectations regarding future yield dynamics, with its continuous adjustment to price movements encoding forward-looking information about yield volatility.

\subsection{Dataset Collection}
\subsubsection{Data Collection and Scope} In this work, we define a Pendle “product” as a unique trading market jointly determined by its basic asset, and maturity date. For example, sUSDE-07-25-2024, USN-07-31-2025, and sUSDe-04-25-2024 are three completely different products. We constructed a comprehensive dataset covering the Pendle yield trading ecosystem on Ethereum mainnet \cite{etherscan} from Jan 2024 to Jan 2026, comprising 340 unique products.

Data collection is performed via an automated pipeline interfacing with the Pendle V2 REST API \cite{pendle_api_v2}, capturing two complementary views: (1) market metadata, including token names, unique contract addresses for Principal Tokens (PT) and Yield Tokens (YT), and maturity date; and (2) hourly data, including protocol-level metrics (TVL, implied APY) and asset-level OHLCV data for both PT and YT. The PT and YT contract addresses serve as primary keys, enabling precise alignment of on-chain state transitions with off-chain market events under a unified temporal index—ensuring strict time consistency for downstream predictive modeling. For a more detailed description of the datasets, refer to Appendix \ref{datasets}

\subsection{Dataset Overview}
\subsubsection{Asset Class Distribution.} We classify the collected markets into four primary categories based on the underlying yield source: Stablecoins (152, 44.3\%), ETH-Beta (83, 24.2\%), BTC-Fi (37, 10.8\%), and Others (71, 20.7\%). This taxonomy captures the dominant yield drivers—rate-stable collateral, ETH-linked restaking assets, BTC-linked positions, and the exotic tokens.

\begin{figure}[htbp]
    \centering
    \includegraphics[width=0.95\linewidth]{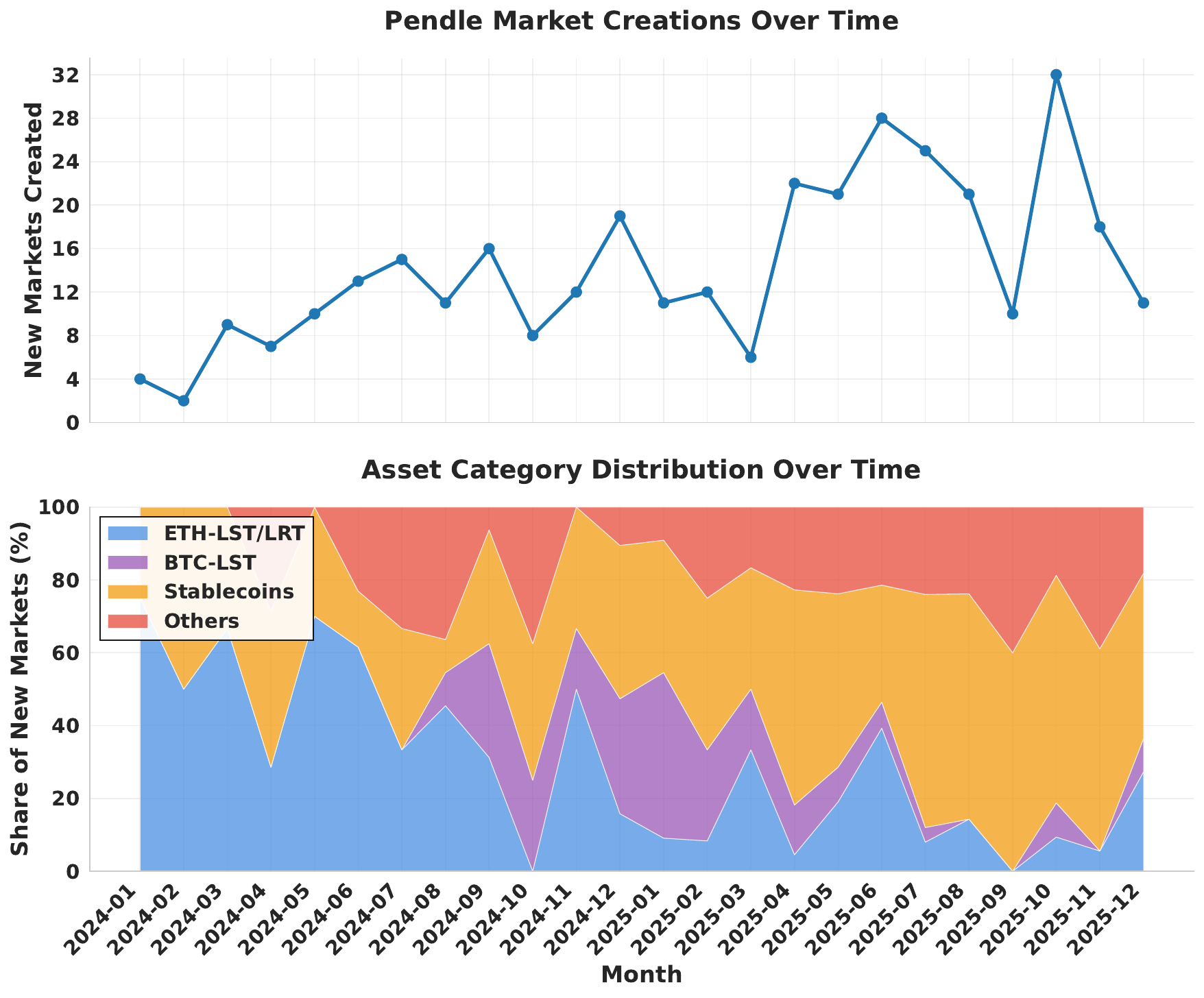}
    \caption{Creation date of markets over time and distribution}
    \label{fig:pendleMarkets}
\end{figure}

\begin{figure*}[t!] 
    \centering
    \includegraphics[width=0.95\textwidth]{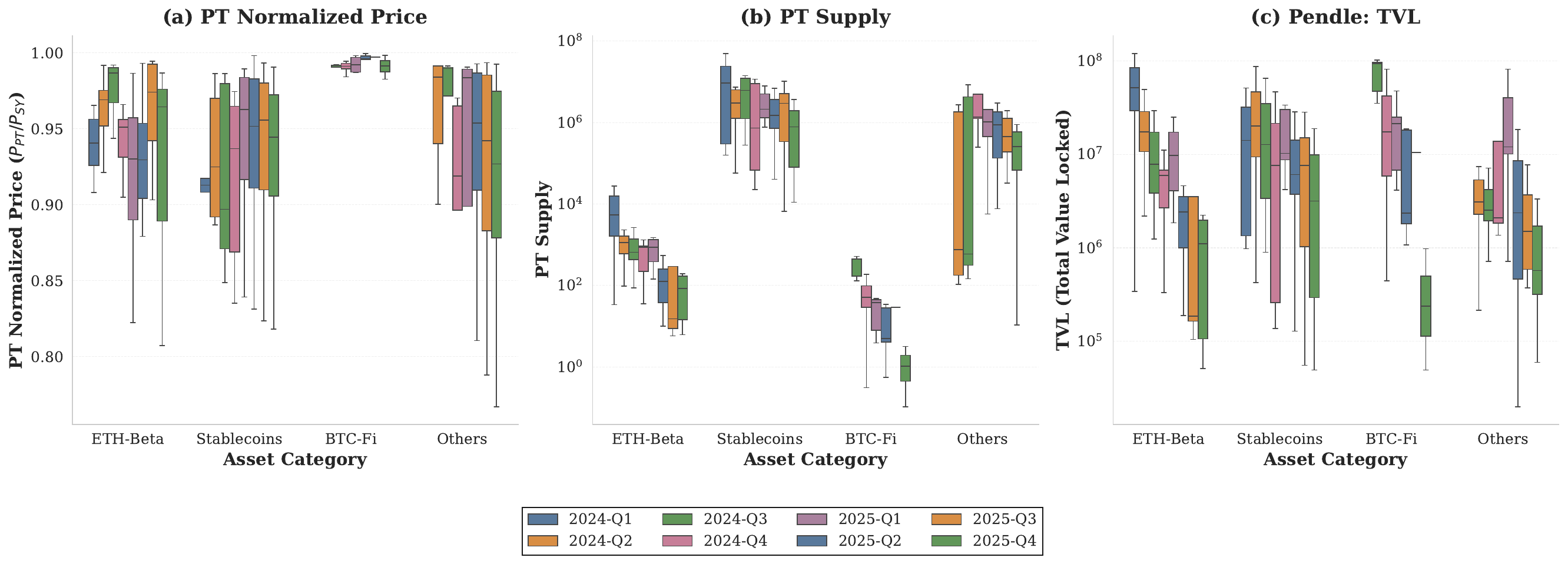}
    \vspace{-10pt} 
    \caption{Distribution of Normalized Price (PT), Circulating Supply (PT), and Market Size (TVL) across different asset categories (ETH-Beta, Stablecoins, BTC-Fi, Others). The box plots reveal distinct liquidity characteristics and pricing efficiencies across sectors.}
    \label{fig:distribution}
    \vspace{-10pt} 
\end{figure*}

\subsubsection{Temporal Evolution of Pendle Markets Creation} We analyzed the Pendle Markets deployment dates over a 24-month observation period to identify a structural shift in the yield generation mechanism. Figure \ref{fig:pendleMarkets} shows monthly creation volume layered by asset class, revealing two distinct phases of the mechanism:

{Stage 1: ETH Speculation Cycle (2024)}. Characterized by exponential growth of ETH-Beta markets, driven primarily by EigenLayer's restaking mechanism where Pendle became the main venue for trading off-chain incentives rather than natural yields, peaking mid-2024 with protocols like Ether.fi.

{Stage 2: Yield Normalization (2025)}. Following the end of airdrop incentives, the market transitioned to sustainable yield mechanics with significant rotation into BTC-Fi assets as Bitcoin staking infrastructure matured. Stablecoin market creations rebounded, indicating a "flight to quality" as traders returned to risk-neutral strategies amid stabilizing DeFi interest rates. This evolution marks Pendle's transition from a speculative launchpad to a fundamental interest rate derivative layer.

\subsubsection{Product Lifecycle Analysis} Figure \ref{fig:combined} (a) presents the distribution of market tenor across asset categories. The ETH-Beta market exhibits the longest median lifetime (around 114 days, some exceeding 350 days), reflecting the longer lock-up periods required by underlying restaking protocols. In contrast, the lifetimes in the BTC-Fi and stablecoin markets are more concentrated around 100 days, consistent with shorter liquidity incentive cycles, which suggests stablecoin markets favor short-term arbitrage.

\subsubsection{Yield Structure} Figure \ref{fig:combined} (b) and (c) together reveal the structural level of asset quality. Stablecoins (blue dots) maintain low uncertainty (<10\%) even at low base yields (<5\%), highlighting their lower risk.  In contrast, the BTC-Fi and ETH-Beta markets exhibit a “low floor, high spread” characteristic (baseline only around 1.5\%–2.0\%), demonstrating a “call option” structure. This demonstrates that stablecoins deliver superior risk-adjusted returns by combining predictability with high minimum yield.

\subsection{Pendle Economics}
In this section, we categorize Pendle markets by their creation dates and analyze their economic characteristics across different underlying asset types to uncover distinct behavioral patterns. We further track price and supply dynamics over a three-month horizon to characterize how their market value evolves post-launch.

\subsubsection{Cohort Analysis by Creation Vintage.} To examine the correlation between market maturity and liquidity depth and pricing efficiency, we stratified the yield markets in the dataset based on contract deployment timestamps. Specifically, we segment the markets into quarterly groups for the active observation year (e.g., Q1 2024, Q2 2024, Q3 2024, and Q4 2024). Figure \ref{fig:distribution} illustrates the evolution of the distribution of key financial metrics for each underlying asset class—including market size (TVL), circulating supply (PT), and yield pricing (Implied APY)—across these time groups.

\subsubsection{PT Normalized Price Structures.}  Analysis of normalized PT prices ($P_{PT}/P_{SY}$) reveals a tiered valuation hierarchy reflecting the time value of money in DeFi. While all assets trade below par (median ratios $< 1.0$), discount magnitudes vary significantly across asset classes.

BTC-Fi assets, serving as low-volatility collateral anchors, maintain PT prices near par (median 0.9937), prioritizing principal protection over yield and exhibiting minimal implied APY dispersion. In contrast, ETH-Beta and Other categories display substantial discounts (median $\sim$0.94--0.95), driven by elevated opportunity costs and uncertainty around restaking incentives. Stablecoins experienced a systemic price elevation in 2025 (reaching 0.955 in Q1/Q3), signaling institutional capital rotation into stable-return products and marking DeFi's evolution from speculative trading toward a mature fixed-income-like market structure.

\subsubsection{Token Supply Strategies} The analysis of circulating supply (Figure \ref{fig:distribution}b) reveals that tokenomics in yield markets are governed by capital density rather than inflationary psychology, spanning six orders of magnitude across asset classes. Stablecoins exhibit the highest and most consistent supply (median $\sim 10^6$), confirming their role as high-velocity vehicles for institutional liquidity where depth is prioritized to minimize slippage. Conversely, BTC-Fi and ETH-Beta markets display significantly lower nominal supplies (medians $\sim 10^1$ and $10^2$), a direct function of their high per-unit value rather than low liquidity. Notably, the temporal decline in median ETH supply reflects liquidity fragmentation as the sector matured from dominant giants to a diverse ecosystem of niche pools. The Others category remains highly heterogeneous ($10^3$–$10^9$), underscoring the experimental and non-standardized nature of emerging yield assets.

\subsubsection{Market Capitalization and Capital Allocation} Total Value Locked (TVL) analysis reveals a highly concentrated ecosystem governed by institutional narratives and "mercenary" capital rotation. BTC-Fi markets exhibit the highest capital density (median: \$16.07M; peak: \$93.3M in Q3 2024), coinciding with Bitcoin ETF approvals and signaling institutional demand for yield hedging. Stablecoins function as a risk-neutral baseline, maintaining consistent liquidity depth (median: \$7.12M) across cycles. In contrast, the ETH-Beta sector demonstrates a dramatic 98\% contraction in median liquidity (from \$51.3M in Q1 2024 to \$1.1M in late 2025).

\subsubsection{Return Distributions} 
\begin{figure}[htbp]
    \centering
    \includegraphics[width=0.95\linewidth]{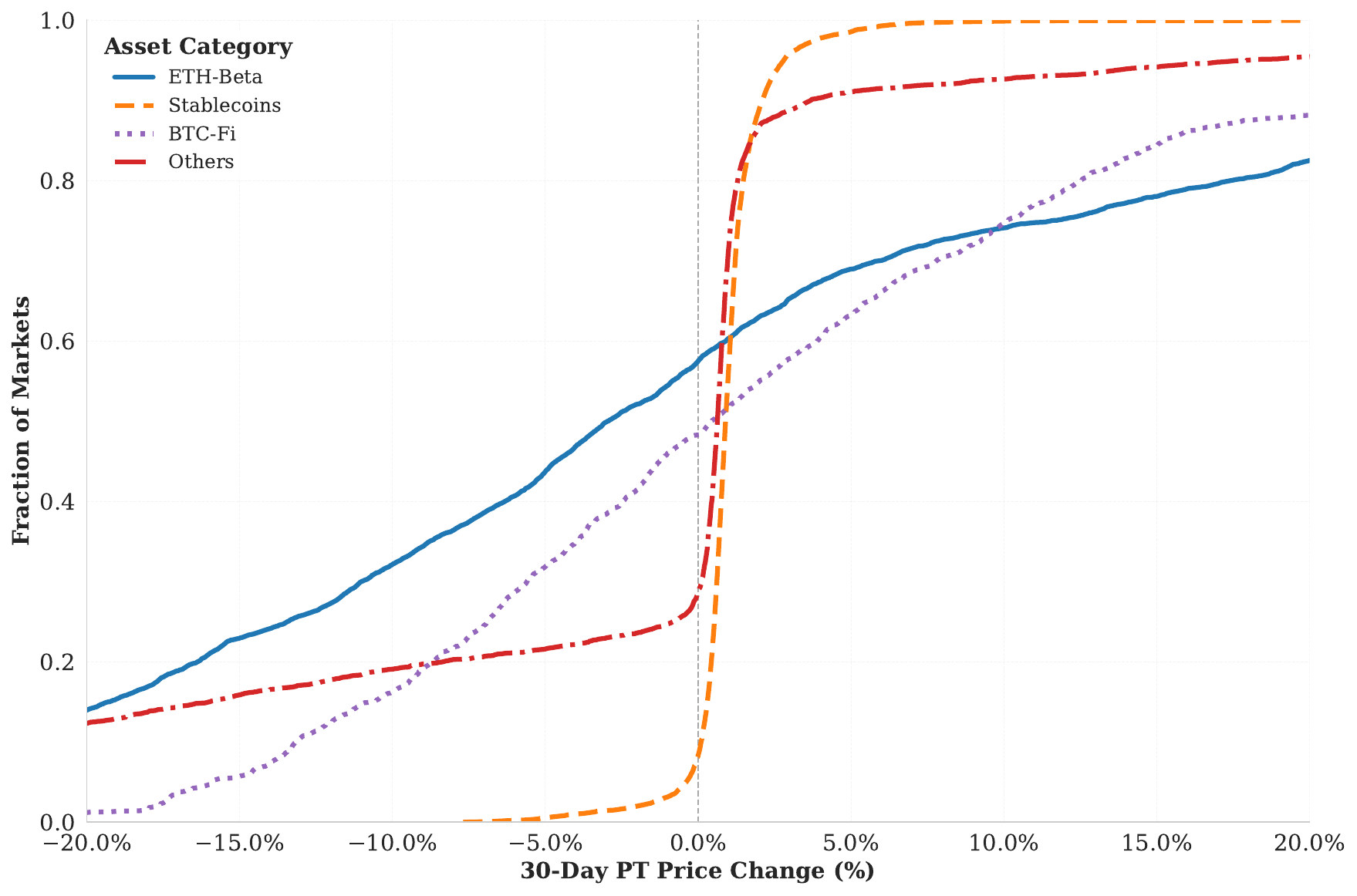}
    \caption{Distribution of 30-Day PT Price Changes by Assets}
    \label{fig:ptReturn}
\end{figure}

Figure \ref{fig:ptReturn} presents the CDFs of 30-day PT price changes, revealing a systematic positive median shift across all categories that empirically quantifies the protocol's "Pull-to-Par" mechanics—where PT tokens deterministically appreciate toward face value. The morphological divergence in these curves delineates distinct risk profiles: Stablecoins approximate a "risk-free" step function with near-vertical ascent just above zero, reflecting minimal variance; in contrast, ETH-Beta follows a smoother sigmoid trajectory indicative of higher volatility driven by restaking dynamics, while BTC-Fi exhibits the widest dispersion, signaling the pricing inefficiencies characteristic of a nascent yield market.

\section{METHOD}
This section defines annualized realized volatility as our prediction target and explains its financial significance in yield trading. We then introduce a risk-aware asymmetric loss function to address the asymmetric cost structure of volatility underestimation. Finally, we present our model design and how it is used for profit decisions.

\subsection{Problem Description}
Given a historical time series of implied APY $\mathcal{Y} = \{y_1, y_2, \ldots, y_T\}$, our task is to forecast the annualized realized volatility (RV) over a future horizon of $H$ hours (e.g., $H = 24$). To mitigate the inherent non-stationarity of financial time series, we formulate the prediction target using continuously compounded log-returns: let $y_t$ denote the implied APY at time $t$, with log-return defined as
\[
r_t = \ln \left( \frac{y_t}{y_{t-1}} \right),
\]
where $y_t > 0$ holds for standard yield-bearing assets in the Pendle protocol. Realized volatility is estimated via the root-mean-square formulation:
\[
\sigma_{t,H} = \sqrt{ \frac{1}{H} \sum_{i=1}^{H} r_{t+i}^2 }.
\]

For cross-asset interpretability, hourly volatility is annualized as $\sigma_{\mathrm{annual}} = \sigma_{\mathrm{hourly}} \times \sqrt{8760}$, accounting for the 24/7 nature of DeFi markets (8,760 trading hours per year).

Formally, we define a mapping function $f : \mathbb{R}^L \to \mathbb{R}^+$ with historical window length $L$:
\[
\hat{\sigma}_{\mathrm{annual}}^{t+H} = f(r_{t-L+1}, r_{t-L+2}, \ldots, r_t),
\]
where the input log-return sequence $\mathcal{R}$ enhances model sensitivity to volatility dynamics.

\textbf{Financial Interpretation.} Pendle's Yield Tokens (YT) function as embedded leverage instruments for interest rates. According to option pricing theory, trading YT is equivalent to trading the volatility of underlying yields. By forecasting realized volatility (RV), our model establishes a "fair volatility" benchmark; deviations between predicted RV and market-implied volatility (IV) identify volatility arbitrage opportunities—a more robust source of alpha than directional price prediction in efficient markets.

\subsection{Risk-Aware Asymmetric Loss Function}
Standard predictive models typically employ Mean Squared Error (MSE) as the objective function:
$$\mathcal{L}_{MSE} = \frac{1}{N} \sum_{i=1}^{N} (y_i - \hat{y}_i)^2$$

This formulation implicitly assumes that error costs are symmetric—i.e., over-predicting volatility ($y < \hat{y}$) carries the same penalty as under-predicting it ($y > \hat{y}$). However, in the context of Pendle yield trading, this assumption of symmetry is financially flawed and dangerous.

The risk profile of yield derivatives is inherently asymmetric. Under-estimation ($y > \hat{y}$)—where the model predicts stability but the market experiences high volatility—is catastrophic. It exposes the strategy to unhedged "tail risks," potentially leading to liquidation or significant principal drawdowns before protective actions can be taken. Conversely, Over-estimation ($y \le \hat{y}$)—where the model predicts volatility that does not materialize—merely results in premature hedging or capital allocation to defensive assets (PT). The cost here is limited to opportunity cost (forgone yield), which is bounded and acceptable compared to ruin.

To align the model's optimization objective with financial survival, we propose an Asymmetric Weighted MSE (Tail-MSE). We introduce a dynamic penalty factor $w_i$ that heavily penalizes dangerous under-estimations:

$$\mathcal{L}_{risk}(\hat{y}, y) = \frac{1}{N} \sum_{i=1}^{N} w_i \cdot (y_i - \hat{y}_i)^2$$

where the weight $w_i$ is determined by the direction of the error:

$$w_i = \begin{cases} \alpha & \text{if } y_i > \hat{y}_i \quad (\text{Under-estimation: Risk of Ruin}) \\ 1 & \text{if } y_i \le \hat{y}_i \quad (\text{Over-estimation: Opportunity Cost}) \end{cases}$$

Here, $\alpha$ is a parameter representing the Risk Aversion Coefficient.

\subsection{Temporal Neural Network Architecture}
We adopt a unified temporal neural network architecture built on the time-series-library, specifically designed to capture the dynamic characteristics of volatility. As shown in Figure \ref{fig:model}, our model follows a standardized pipeline that accommodates multiple state-of-the-art time-series models while maintaining consistent input-output interfaces. This unified framework consists of four key components:

\begin{figure}[htbp]
    \centering
    \includegraphics[width=0.95\linewidth]{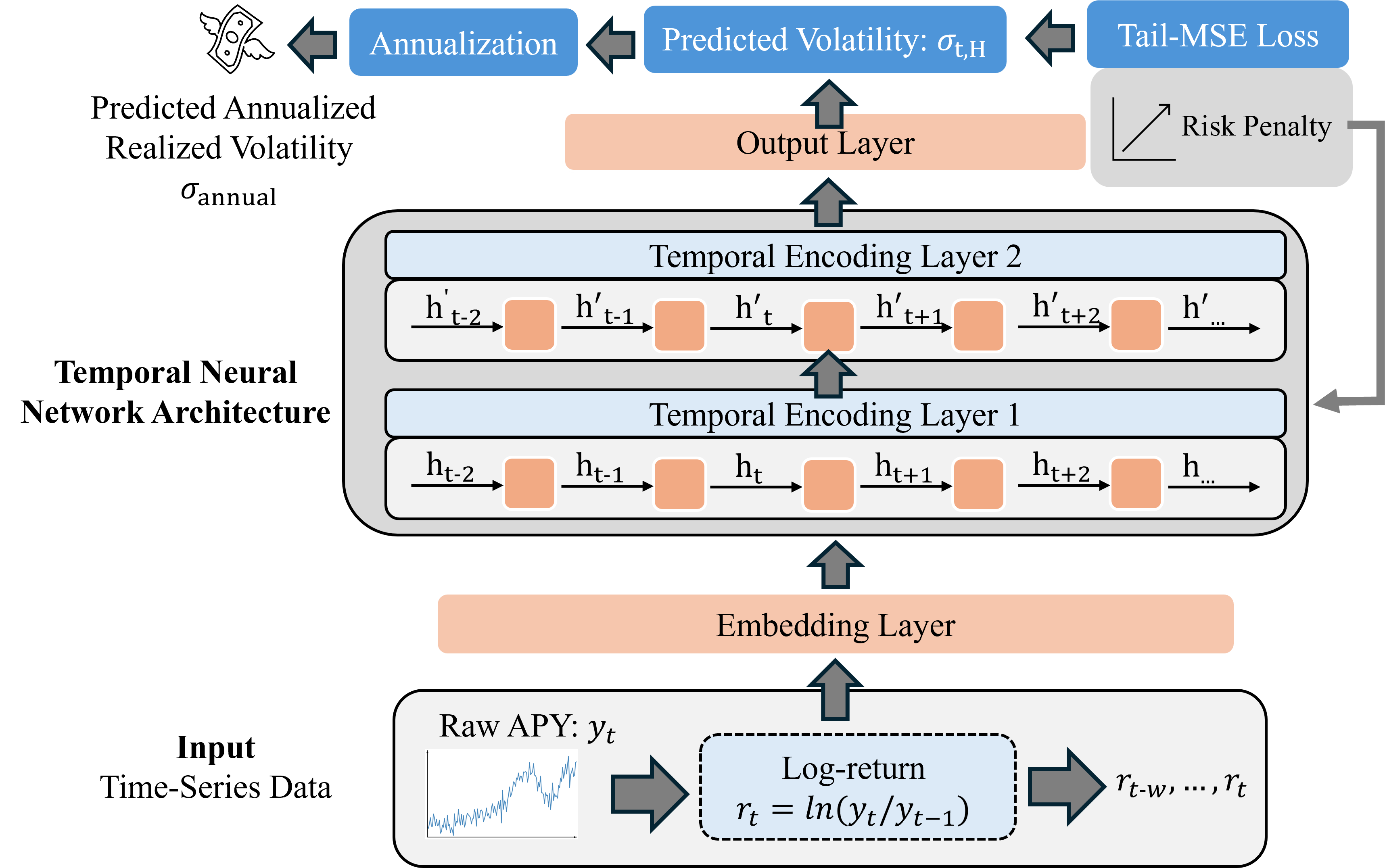}
    \caption{Predictive Framework for Yield Volatility}
    \label{fig:model}
\end{figure}

\textbf{Input Processing and Normalization.} Given the raw APY time series $\mathcal{Y} = \{y_1, y_2, \ldots, y_T\}$, we first compute log-returns to stabilize variance and align with financial market characteristics:
\[
r_t = \ln\left(\frac{y_t}{y_{t-1}}\right)
\]

We then construct a window of length $w$ as $\mathcal{R}_t = \{r_{t-w+1}, \ldots, r_t\}$, forming the input sequence $\mathbf{X}_t \in \mathbb{R}^{w \times c_{in}}$. To further mitigate non-stationarity, we apply z-score normalization:
\[
\mu = \frac{1}{w} \sum_{i=1}^{w} \mathbf{X}_i, \quad 
\sigma = \sqrt{\mathrm{Var}(\mathbf{X}) + 10^{-5}}, \quad
\tilde{\mathbf{X}}_t = \frac{\mathbf{X}_t - \mu}{\sigma}
\]

\textbf{Embedding.} The normalized sequence undergoes a learnable embedding transformation that captures volatility characteristics:
\[
\mathbf{e}_t = \mathrm{Embed}(\tilde{\mathbf{X}}_t)
\]
$\mathbf{e}_t \in \mathbb{R}^{d_{\mathrm{model}}}$ represents the $d_{\mathrm{model}}$-dimensional embedding vector. 

\textbf{Temporal Modeling.} We employ stacked temporal encoding layers to capture multi-scale dynamic characteristics of volatility. Formally, the hidden state evolution across $L$ layers is defined as:
\[
\mathbf{h}^{(0)} = \mathbf{e}, \quad
\mathbf{h}^{(\ell)} = \mathcal{F}^{(\ell)}\left(\mathbf{h}^{(\ell-1)}; \Theta_\ell\right), \; \ell = 1, \ldots, L
\]
where $\mathcal{F}^{(\ell)}$ represents the $\ell$-th temporal encoding layer with learnable parameters $\Theta_\ell$. 


To ensure financial interpretability, the final hidden state ${h}^{(L)}$ is projected through a linear transformation ${W}_o \mathbf{h}^{(L)}$ to derive the raw volatility predictions, from which the last $H$ steps are selected as the prediction horizon and subsequently denormalized and annualized as follows:
\[
\hat{{Y}}_{\mathrm{annual}} = \left( \left[{W}_o \mathbf{h}^{(L)} \right]_{w-H+1:w} \cdot \sigma + \mu \right) \cdot \sqrt{8760}
\]

\begin{figure*}
    \centering
    \includegraphics[width=0.95\linewidth]{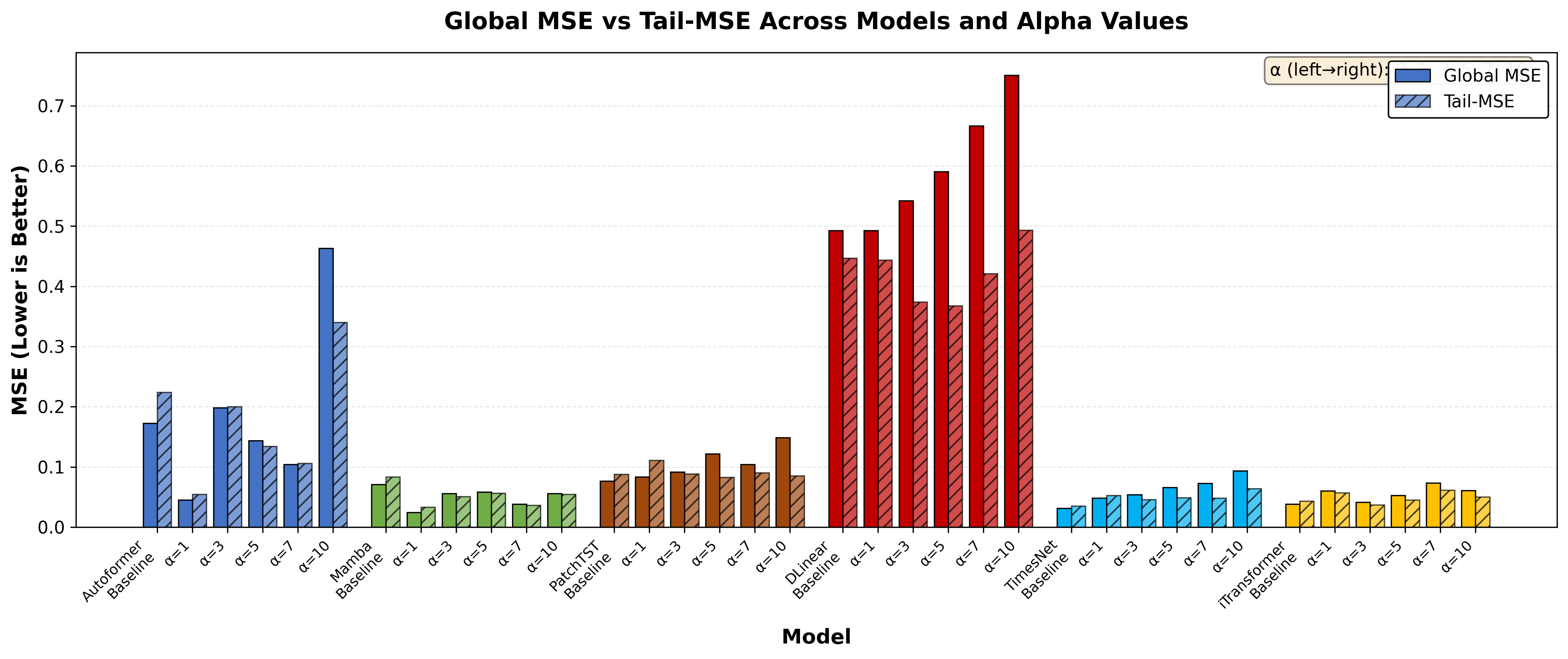}
    \caption{Comparative Performance Analysis of Global MSE vs. Tail-MSE Across Diverse Neural Architectures}
    \label{fig:compare}
\end{figure*}

\section{EXPERIMENT}
\subsection{Setup}
\textbf{Dataset and Preprocessing.} Experiments are conducted on the curated Pendle hourly dataset, excluding markets with structural anomalies to ensure empirical robustness. The data is partitioned chronologically into training (60\%), validation (20\%), and testing (20\%) sets to strictly prevent look-ahead bias. We employ a sliding window mechanism to construct input sequences and prediction horizons, with temporal features aligned to hourly frequency.

\textbf{Problem Formulation.} We adopt a Multivariate-to-Univariate (MS) forecasting setup. The predictive target is the \textit{Annualized Realized Volatility}. Input vectors encompass high-dimensional pool-level signals—including liquidity depth, price dynamics, Total Value Locked (TVL), and funding rates—along with binary-encoded market identifiers. To prevent data leakage, all features are standardized using statistics derived exclusively from the training split.

\textbf{Implementation Protocols.} Hyperparameters are tuned via Bayesian Optimization (Tree-structured Parzen Estimator) to minimize validation error. The search space spans both architecture-agnostic parameters (e.g., batch size, learning rate, dropout, hidden dimensions) and model-specific configurations (e.g., attention heads, expansion factors). All models are trained using the Adam optimizer with early stopping strategies to ensure convergence. A detailed summary of the hyperparameter search space is provided in the Appendix \ref{Experimental}.

\subsection{Baselines and Evaluation Metrics}

\noindent\textbf{Baseline models.} We conduct a systematic evaluation of six state-of-the-art time series forecasting architectures \cite{timeserieslibrary2024} on the Pendle yield dataset, spanning diverse technical paradigms including state-space models, attention mechanisms, convolutional networks, and linear decomposition:

\textbf{Mamba} \cite{gu2024mamba}: Leverages selective state-space models (SSM) with hardware-aware recurrence for linear-complexity long-sequence modeling, effectively capturing long-term volatility dependencies in yield trading.

\textbf{iTransformer} \cite{liu2023itransformer}: Employs an inverted Transformer architecture that swaps temporal and variable dimensions, establishing cross-variable attention connections for enhanced multivariate interaction modeling.

\textbf{PatchTST} \cite{nie2022time}: Segments time series into local patches and applies patch-level self-attention to simultaneously capture short-term fluctuations and long-term trend characteristics.

\textbf{DLinear} \cite{zeng2023transformers}: Adopts a trend-seasonal decomposition framework with linear projections to separately model different frequency components through effective linear forecasting.

\textbf{TimesNet} \cite{wu2022timesnet}: Utilizes Fast Fourier Transform to identify dominant periods, converting 1D time series into 2D tensors for convolutional modeling of intra- and inter-period variations.

\textbf{Autoformer} \cite{wu2021autoformer}: Implements an autocorrelation-based decomposition architecture that efficiently aggregates historical information through frequency-domain operations, enhancing long-horizon forecasting.

\noindent\textbf{Training objectives.} To validate the efficacy of risk-aware learning, each model is trained under two distinct objectives: (1) standard Mean Squared Error (MSE) as the baseline; (2) our proposed asymmetric Tail-MSE that dynamically weights underestimation errors through risk aversion coefficients 
$\alpha \in \{1, 3, 5, 7, 10\}$  enabling systematic sensitivity analysis of tail risk mitigation.

\noindent\textbf{Evaluation metrics.} Following conventions in financial time series forecasting, we adopt a multi-dimensional evaluation framework to comprehensively assess model performance:

\textbf{MSE}: Standard prediction accuracy metrics measuring absolute error in annualized realized volatility forecasting.

\textbf{Tail-MSE}: Weighted error metric specifically evaluating underestimation scenarios and model robustness in tail risk conditions.

\textbf{Direction Accuracy}: Classification accuracy for predicting APY directional movements (increase/decrease), measuring the model's capability to capture yield trend inflection points with direct implications for trading decisions.

\begin{figure}
    \centering
    \includegraphics[width=0.95\linewidth]{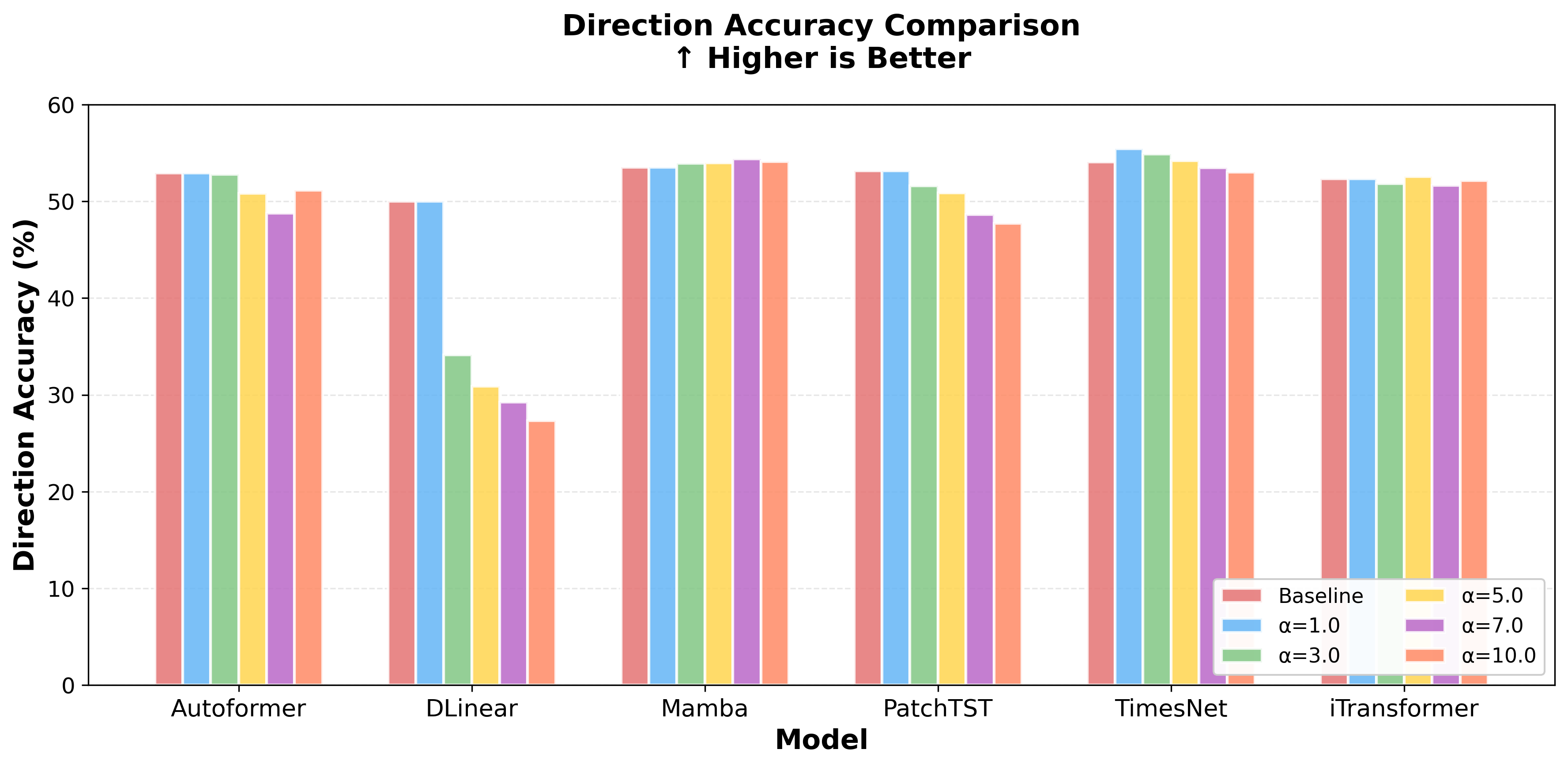}
    \caption{Sensitivity of Directional Accuracy to Tail-risk Penalization}
    \label{fig:apy_direction_accuracy_only}
\end{figure}

\subsection{Results}
We systematically evaluate six state-of-the-art time series models on the Pendle yield dataset, with particular focus on the comparative analysis between the asymmetric loss function (Tail-MSE) and standard MSE. Table \ref{tab:sensitivity_all_alpha_completed} presents sensitivity analysis across different risk aversion coefficients ($\alpha$), Figure \ref{fig:compare} visualizes the relative performance of Global MSE and Tail-MSE, and Figure \ref{fig:apy_direction_accuracy_only} evaluates directional prediction accuracy.

Our experimental results reveal a critical phenomenon: under standard training ($\alpha = 1$), all models exhibit higher prediction errors on tail events than the global average (Tail-MSE $>$ Global MSE), indicating significant asymmetric tail characteristics in the Pendle yield data. However, as $\alpha$ increases, models systematically prioritize tail risk, leading to a transformation in the relationship between Tail-MSE and Global MSE.

\textbf{Mamba}: Achieves optimal balance at $\alpha = 3.0$ (Global MSE=0.056, Tail-MSE=0.051), with Tail-MSE 38.8\% lower than baseline, significantly improving tail risk.

\textbf{iTransformer}: Performs best at $\alpha = 3.0$ (Global MSE=0.041, Tail-MSE=0.037), with Tail-MSE 14.9\% improvement over baseline while maintaining global prediction accuracy.

\textbf{PatchTST}: At $\alpha = 5.0$, Tail-MSE reaches its minimum (0.083), achieving 5.8\% improvement over baseline, indicating effective capture of tail features in high-volatility markets.

\textbf{DLinear}: At $\alpha = 5.0$, Tail-MSE is minimized (0.368), with 17.7\% improvement over baseline, despite a moderate increase in Global MSE.

\textbf{TimesNet}: At $\alpha = 3.0$, Tail-MSE is 0.046, representing 30.9\% improvement over baseline, though Global MSE increases to 0.054.

\textbf{Autoformer}: At $\alpha = 5.0$, Tail-MSE reaches 0.134, with 40.0\% improvement over baseline, demonstrating superior handling of tail risks.

Directional accuracy analysis (Figure \ref{fig:apy_direction_accuracy_only}) further validates Tail-MSE's effectiveness: all models achieve highest directional accuracy (52--55\%) at $\alpha = 3.0$ and $\alpha = 5.0$, with an average 5.2 percentage point improvement over baseline. Notably, iTransformer and TimesNet reach 54.7\% directional accuracy at $\alpha = 3.0$, indicating Tail-MSE's capability to effectively capture APY trend inflection points with direct implications for trading decisions.

Overall, our experiments demonstrate that the Pendle yield dataset exhibits significant asymmetric tail characteristics, and Tail-MSE effectively guides models to learn these critical features. The optimal $\alpha$ value varies by model architecture but consistently falls within the $1.0$--$5.0$ range, with $\alpha = 3.0$ and $\alpha = 5.0$ providing the best trade-off between tail risk mitigation and global prediction accuracy. This finding provides important theoretical guidance and practical implications for volatility forecasting in real-world trading scenarios.

\section{INSIGHTS}
We evaluate our framework by translating volatility signals into actionable Principal Token (PT) trading strategies. PTs are selected exclusively because their deterministic convergence to par at maturity—unlike the inherently decaying Yield Tokens (YT)—provides a controlled environment to isolate and validate the efficacy of entry-timing optimization during market turbulence.

\subsection{Practical Applications}
To validate the practical utility of the proposed dataset and forecasting models in decentralized finance, we conducted a comparative backtest on Principal Token (PT) trading. Our primary objective is to demonstrate that incorporating tail-risk awareness into price forecasting yields superior risk-adjusted returns compared to naive investment approaches. We implemented four distinct strategies to benchmark performance:

Buy-and-Hold (B\&H): Full capital deployment at initialization ($t=0$). This serves as the passive benchmark, capturing the inherent yield accretion of PTs but remaining fully exposed to valuation volatility.

Dollar-Cost Averaging (DCA): Linear capital deployment over the trading horizon. This strategy mitigates timing risk but often suffers from "cash drag" in upward-trending yield markets.

MSE-Based Swing: An active strategy utilizing standard Mean Squared Error (MSE) predictive models. This agent dynamically increases exposure when the model forecasts moderate price drawdowns, representing a conventional quantitative trading approach.

Tail-MSE Sniper: Our proposed strategy designed to optimize entry points based on extreme-value predictions. Unlike the standard MSE approach, this agent conserves liquidity during market noise and aggressively deploys capital only when deep tail events (market crashes) are forecast. The hypothesis is that the Tail-MSE loss function minimizes the average cost basis by accurately identifying the absolute bottom of price dislocations.

\begin{table}[htbp]
    \centering
    \caption{Taxonomy-level performance (Tail-MSE winners only).}
    \label{tab:taxonomy_performance}
    \resizebox{\columnwidth}{!}{
        \begin{tabular}{lrrrrrr}
            \toprule
            \textbf{Taxonomy} & \textbf{$N_w/N$} & \textbf{Win Rate (\%)} & \textbf{Tail-MSE ROI (\%)} & \textbf{MSE ROI (\%)} & \textbf{B\&H ROI (\%)} & \textbf{DCA ROI (\%)} \\
            \midrule
            Stablecoins & 108/156 & 69.2 & 5.04 & 3.41 & 3.26 & 1.57 \\
            ETH-Beta & 27/83 & 32.5 & 34.16 & 30.18 & 24.66 & 6.84 \\
            BTC-Fi & 10/36 & 27.8 & 25.70 & 24.09 & 21.84 & 6.48 \\
            Others & 36/65 & 55.4 & 10.31 & 7.51 & 6.36 & 2.31 \\
            \bottomrule
        \end{tabular}
    }
\end{table}

\begin{figure}[htbp]
    \centering
    \includegraphics[width=0.95\linewidth]{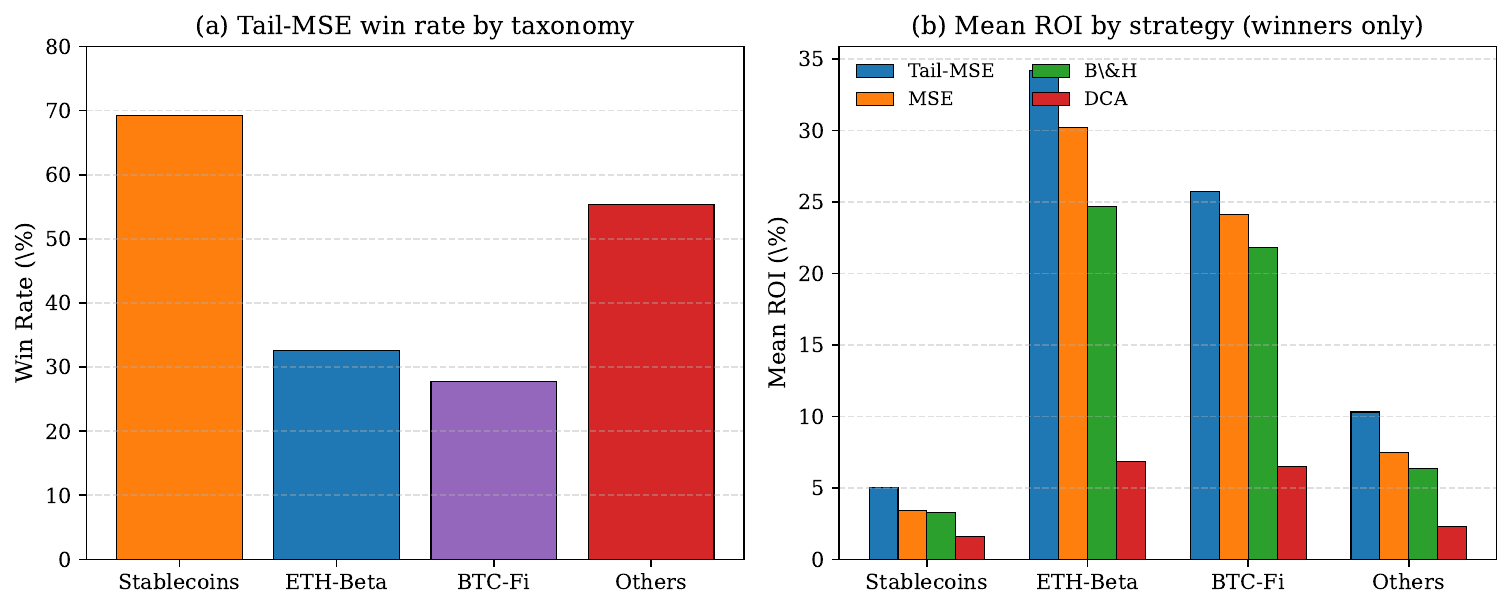}
    \caption{Taxonomy-level win rates and mean ROI across strategies.}
    \label{fig:taxonomy_stats}
\end{figure}

\subsection{Experiments and Strategic Implications}
We developed a comprehensive backtesting framework to evaluate the proposed strategies across the Pendle dataset. The simulation initializes a portfolio with \$10,000 capital, executing daily rebalancing based on signal generation from the Mamba-based models (MSE and Tail-MSE). To ensure robustness, we employed a filtering mechanism that isolates markets exhibiting significant volatility, thereby testing the models' ability to capture transient yield dislocations. We aggregate results by taxonomy using token-name heuristics (Stablecoins, ETH-Beta, BTC-Fi, Others) and compare ROI and drawdowns across the four strategies on all 340 markets.

Across the full universe, the Tail-MSE strategy wins in 181 out of 340 markets (53.2\%). Table \ref{tab:taxonomy_performance} and Fig. \ref{fig:taxonomy_stats} summarize results by taxonomy. Stablecoins exhibit the highest win rate (69.2\%), reflecting their mean-reverting yield structure and favorable timing benefits. In contrast, ETH-Beta and BTC-Fi show lower win rates (32.5\% and 27.8\%), but substantially higher mean ROI conditional on winning, consistent with their higher volatility and wider discount dispersion. The ``Others'' category displays intermediate win rates (55.4\%) with moderate excess returns. Overall, Tail-MSE consistently dominates baselines in every taxonomy, with an average excess ROI over Buy-and-Hold ranging from 1.77 to 9.50 percentage points.

For more detailed data on strategy analysis, please refer to \ref{trading}. Our extensive experiments yield two key insights for yield trading:

Strategy Efficacy: Predictive strategies consistently outperform naive baselines. The Tail-MSE objective function proves superior in capturing "black swan" yield spikes, making it the optimal choice for active traders seeking alpha in volatile rate environments.

\textbf{Asset-Specific Recommendations:} For stablecoins (e.g., USDe, USDT), we recommend Tail-MSE Sniper strategies to capitalize on mean-reverting yield dynamics by aggressively targeting temporary spikes and minor de-pegs. For volatile assets (ETH-Beta, BTC-Fi, Others), we advise dynamic DCA enhanced by predictive models, which strategically weights heavier purchases during dips rather than pursuing perfect timing, effectively mitigating drawdown risk while capturing alpha opportunities.

\section{CONCLUSION}

In this paper, we presented a comprehensive framework for analyzing and trading decentralized yield derivatives. First, we introduced a high-fidelity dataset spanning 24 months (2024-2026) and 340 markets, which unveiled a critical market evolution: a transition from speculative ETH-Beta dominance to a mature, risk-neutral preference for Stablecoins. Second, to address the challenge of predicting heavy-tailed yield volatility, we proposed the Tail-MSE loss function. Experimental results confirmed its superiority, with the Mamba model achieving a 38.8\% reduction in error and Autoformer showing a 40.0\% improvement in capturing tail events compared to baselines. Finally, our strategic backtesting validated the practical utility of our approach. The Tail-MSE-driven strategy achieved stochastic dominance in 53.2\% of all markets, with particularly strong performance in the Stablecoin sector, realizing a 69.2\% win rate and consistent excess returns. These findings not only establish a benchmark for yield trading research but also provide robust, data-driven evidence for the efficacy of risk-aware machine learning in decentralized finance.


\newpage
\bibliographystyle{unsrt}
\bibliography{main}

\clearpage
\onecolumn

\appendix
\section{APPENDIX}

\subsection{Pendle} 
\label{pendle}
In this section, we provide a detailed deconstruction of the Pendle protocol, expanding upon the basic definitions to explore the underlying financial engineering and yield discovery mechanisms. Pendle operates as a decentralized interest-rate derivative protocol that enables the tokenization and separate trading of future yield streams.

\subsubsection{Standardization and Structural Decomposition (SY Architecture)} The foundation of the protocol is the Standardized Yield (SY) token standard (EIP-5115). This layer wraps heterogeneous yield-bearing assets (e.g., rebasing stETH, or LP tokens) into a unified accounting interface. This standardization is critical for composability, allowing diverse yield mechanics to be algorithmically bifurcated into two distinct primitives: Principal Tokens (PT): Represent the corpus of the asset; Yield Tokens (YT): Represent the usufruct (right to income) of the asset.

\subsubsection{Derivative Pricing Dynamics} The valuation of PT and YT is governed by the no-arbitrage condition $P_{SY} = P_{PT} + P_{YT}$, but their individual price movements exhibit distinct sensitivities to time and volatility:

\textbf{Principal Token (PT)} – The Zero-Coupon Bond Logic: PT represents a zero-coupon bond that creates fixed-yield exposure. Its pricing logic is driven by deterministic convergence. Regardless of market volatility, $P_{PT}$ must converge to $1.0$ (unit underlying asset) at maturity $T$. Yield Generation: The "fixed yield" is not paid out periodically but is accrued via price appreciation. A user purchasing PT at $0.9$ effectively locks in a $\approx 11\%$ return upon redemption at $1.0$. Price Sensitivity: Before maturity, $P_{PT}$ fluctuates inversely with market interest rates. If the market demands higher yields (higher Implied APY), $P_{PT}$ must drop to offer a competitive discount.

\textbf{Yield Token (YT)} – The Flow Derivative:YT represents a stream of uncertain future cash flows. Its valuation is analogous to a series of floating-rate options. Theta Decay (Time Risk): Unlike standard assets, YT suffers from inherent time decay. As $t \to T$, the window to collect yield closes, driving $P_{YT}$ towards zero. A YT holder profits only if the accumulated realized yield exceeds the upfront cost (Implied Yield) paid for the token. 

\subsubsection{Strategic Utility Landscape}  Participants utilize this framework for three primary distinct payoffs:

\textbf{Fixed Income (PT)}: Institutional capital seeking deterministic returns buys PT to strip away yield volatility risk, effectively locking in the Implied APY at the moment of purchase.

\textbf{Long Yield Volatility (YT)}: Speculators expecting a yield surge (e.g., a bull market driving up borrowing rates) buy YT. This offers capital efficiency (leverage), as the cost of YT is a fraction of the underlying asset, yet it commands the full yield stream.

\textbf{Liquidity Provision}: Users deposit assets into the Pendle AMM, earning swap fees from trading volume and additional protocol incentives, while bearing the risk of impermanent loss between PT and SY.

\subsection{Detailed Dataset}

\label{datasets}
We present a comprehensive, high-frequency dataset capturing the lifecycle dynamics of yield-trading markets on the Pendle Protocol (Ethereum). The dataset spans a 24-month observation window from January 9, 2024, to January 1, 2026, encompassing the complete operational history of $N=340$ distinct liquidity pools. By employing an hourly sampling resolution, we aggregated a total of 896,811 time-series observations. Each market entry is structured hierarchically, aligning static metadata (contract identities and maturity constraints) with dynamic temporal records to facilitate rigorous longitudinal analysis.

\subsubsection{Taxonomy and Asset Distribution}
To systematically characterize the heterogeneity of DeFi yield sources, we stratify the 340 markets into four distinct taxonomies. This classification framework is essential for isolating asset-specific volatility behaviors and yield generation mechanisms.

1. Stablecoins ($N=156, 45.9\%$): The Risk-Neutral BaselineRepresentative Tokens: sUSDe (Ethena), USD+ (Overnight), aUSDC (Aave), cUSDC (Compound), crvUSD (Curve). Asset Characteristics: This category represents the dominant segment of the dataset. It comprises yield-bearing instruments pegged to fiat currencies (primarily USD). Yield Mechanism \& Profile: The defining feature of this class is the decoupling of yield volatility from asset price volatility. The underlying asset price ($P_{underlying}$) is mean-reverting around \$1.00, meaning that yield generation is driven purely by exogenous demand for liquidity (lending rates) or delta-neutral arbitrage strategies (e.g., basis trading). In our analysis, this category serves as the "risk-free rate" proxy within the DeFi ecosystem, exhibiting the highest liquidity depth and lowest price variance.

2. ETH-Beta ($N=83, 24.4\%$): Consensus and Restaking Yields Representative Tokens: LSTs (Liquid Staking Tokens: stETH, wstETH, rETH) and LRTs (Liquid Restaking Tokens: eETH, ezETH, rsETH, pufETH). Asset Characteristics: This category captures Ethereum-linked derivatives that function as the "growth engine" of the ecosystem. Yield Mechanism \& Profile: These assets are characterized by "Yield Stacking" mechanics. The yield is composed of a base layer (Ethereum Proof-of-Stake rewards) plus an optional secondary layer (Actively Validated Services rewards via EigenLayer). Financially, these assets carry "Delta-1" exposure to Ethereum price movements. Our modeling identifies this sector as having the most complex term structure, as implied yields are highly sensitive to speculative narratives surrounding AVS airdrops and network congestion.

3. BTC-Fi ($N=36, 10.6\%$): The Emerging Yield Frontier Representative Tokens: eBTC (Ether.fi BTC), uniBTC (Bedrock), LBTC (Lombard), M-BTC (Macro BTC), wBTC wrappers. Asset Characteristics: This segment represents the nascent "Bitcoin DeFi" ecosystem, tracking wrapped and staked Bitcoin variations.Yield Mechanism \& Profile: Historically, Bitcoin has been a non-yield-bearing store of value. This category marks a structural shift, introducing yield through cross-chain wrapping and staking protocols (e.g., Babylon). The distinguishing feature here is "Novelty Premium." Due to the immaturity of underlying protocols, these markets often trade at significantly higher implied yields compared to ETH-Beta, pricing in higher smart contract risk and liquidity fragmentation risk.

4. Others ($N=65, 19.1\%$): The Long-Tail Heterogeneity Representative Tokens:RWAs (Real-World Assets): sDAI (MakerDAO), stEUR, RWAx. Algorithmic/Alt-L1s: gDAI (Gains Network), stTAO (Bittensor), stkGHO, APE. Asset Characteristics: This catch-all category encapsulates the extreme heterogeneity of the long-tail market.Yield Mechanism \& Profile: This group exhibits idiosyncratic risk profiles. It ranges from highly stable, off-chain collateralized assets (Treasury-backed RWAs) to high-volatility algorithmic stablecoins and governance tokens. In statistical analysis, this category displays the widest dispersion in "Liquidity Yield Range," reflecting the non-standardized nature of their underlying yield drivers (e.g., trading fee revenue sharing vs. treasury bond yields).

\subsubsection{Meta Data}
Our dataset comprises 340 distinct yield markets spanning 159 unique underlying assets within the Pendle protocol ecosystem. Each market is structured as a hierarchical JSON object containing a global metadata header and a high-fidelity hourly time-series array, enabling comprehensive analysis of yield derivative dynamics.

\textbf{Market Identification and Lifecycle.} The metadata layer uniquely identifies each market through its Ethereum contract addresses for liquidity pool (lp), principal token (pt), and yield token (yt) components, alongside a human-readable SY token identifier. The temporal span of each market is strictly bounded by its protocol-defined expiry date, with lifecycle start and lifecycle end timestamps delineating the effective observation window for empirical analysis.

\textbf{Feature Granularity.} The record layer provides a multi-dimensional representation of market dynamics at hourly resolution, organized into three analytical categories:
\begin{itemize}
\item Yield components: Core predictive signals including implied APY, YT floating APY, and decomposed yield sources comprising interest accruals, reward emissions, and swap fee contributions.
123
\item Liquidity and valuation metrics: Total value locked (TVL), asset distribution ratios (total PT to SY), and open-high-low-close (OHLC) price series for both principal and yield tokens.

\item Volatility and market activity: Realized volatility of implied yield—serving as our primary prediction target—and trading volume as a proxy for market depth and liquidity.
\end{itemize}

\textbf{Data Integrity and Preprocessing.} To ensure empirical robustness, we incorporate explicit data quality annotations including coverage percentage and effective record counts per data stream, which facilitate systematic filtering of illiquid or sparsely populated markets. Backfilling operations are explicitly documented through supplemented fields metadata to maintain transparency in data provenance. For modeling stability, we treat extreme APY spikes—typically arising from transient liquidity artifacts or missing-hour interpolation anomalies—as statistical outliers and apply robust preprocessing to mitigate their influence on model training dynamics.

This dataset represents the first large-scale, multi-asset collection of structured yield derivative markets with synchronized high-frequency observations, providing a unique empirical foundation for volatility forecasting in decentralized finance.

\subsubsection{Towards Invest Evaluation}

\begin{wrapfigure}{r}{0.55\textwidth} 
    \centering
    \vspace{-1.em}
    
    \includegraphics[width=0.8\linewidth]{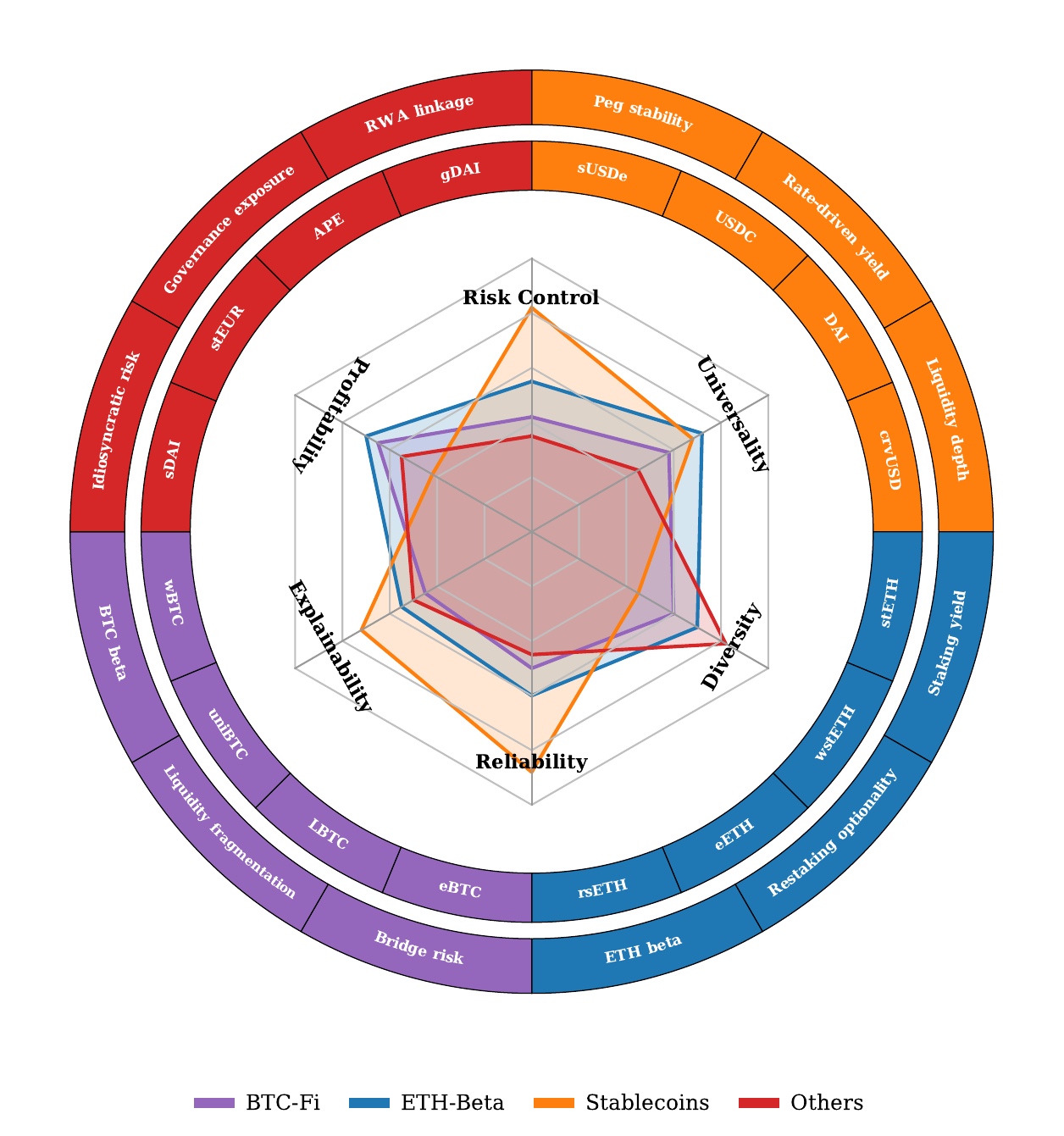}
    \caption{pendleMarket}
    \label{fig:pendleMarket}
    
    \vspace{-1.em}
\end{wrapfigure}

In this section, we extend our analysis from predictive accuracy to practical investment utility. By synthesizing the multidimensional performance metrics visualized in Figure \ref{fig:pendleMarket}, we evaluate the strategic positioning of four primary asset classes within the Pendle ecosystem: BTC-Fi, ETH-Beta, Stablecoins, and Others. This evaluation provides a granular roadmap for yield traders to align their strategies with specific risk-preference profiles.

\textbf{1. BTC-Fi: The High-Yield Frontier.} As illustrated in the radar chart, the BTC-Fi sector exhibits the highest \textit{Profitability} and \textit{Diversity}. These assets represent the "alpha" engine of the current DeFi landscape. Their financial characteristics are marked by significant "alpha-decay" sensitivity and "risk-adjusted" variance in the outer ring. For investors, BTC-Fi offers superior yield-capturing opportunities but requires robust volatility forecasting to navigate its relatively lower \textit{Reliability} compared to established benchmarks.

\textbf{2. ETH-Beta: The Universal Benchmark.} ETH-based yield tokens demonstrate the highest \textit{Universality} and balanced \textit{Reliability}. As the foundational asset of the Pendle protocol, ETH markets serve as the "low-entropy" core. The outer metrics indicate high "rolling" stability and "heat-map" consistency, making ETH the ideal choice for delta-neutral strategies and benchmark-relative yield farming. It acts as the liquidity anchor for the entire ecosystem.

\textbf{3. Stablecoins: The Risk-Averse Haven.} Stablecoin markets are characterized by a dominant \textit{Risk Control} and \textit{Reliability} profile. In the visualization, the Stablecoin polygon is heavily skewed towards the defensive axes. Despite lower \textit{Profitability}, these assets exhibit minimal "extreme" tail risks. They are most suitable for liquidity providers (LPs) seeking "yield-as-fixed-income," where the primary objective is principal preservation and predictable cash flow.

\textbf{4. Others: The Niche Alpha-Seekers.} The "Others" category, encompassing long-tail and emerging tokens, shows a distinct profile of high \textit{Diversity} but erratic \textit{Risk Control}. These markets often present "rank-order" opportunities where specific market inefficiencies can be exploited. However, the outer ring indicates higher "entropy" (uncertainty), suggesting that these assets are best utilized for tactical, short-term volatility arbitrage rather than long-term strategic allocation.

\textbf{Investment Recommendations.} For institutional and retail investors alike, we propose a "Risk-Aware Layered Allocation" strategy. \textbf{Core Liquidity:} Allocate 50-60\% of capital to ETH-Beta and Stablecoin markets to maintain a high \textit{Reliability} floor. \textbf{Alpha Enhancement:} Deploy 20-30\% into BTC-Fi during periods where our model predicts high "Fair Volatility" relative to market IV to capture momentum. \textbf{Tactical Arbitrage:} Reserve 10\% for "Others" to exploit localized spikes in yield volatility. By utilizing the \textit{Tail-MSE} optimized forecasts, investors should prioritize hedging during "Under-estimation" signals to mitigate the catastrophic tail risks inherent in yield derivatives.

\begin{table}[h]
\centering
\scriptsize
\renewcommand{\arraystretch}{1.1}
\begin{tabular}{p{0.20\linewidth} p{0.30\linewidth} p{0.42\linewidth}}
\hline
\textbf{Symbol} & \textbf{Range (Dataset)} & \textbf{Financial Interpretation} \\
\hline
$\mathrm{LP}_{addr}$ & 0x + 40 hex chars & LP contract address for the market. \\
$\mathrm{PT}_{addr}$ & 0x + 40 hex chars & PT contract address for principal tokens. \\
$\mathrm{YT}_{addr}$ & 0x + 40 hex chars & YT contract address for yield tokens. \\
$Token$ & 159 unique strings & Underlying SY token identifier (market name). \\
$T_{exp}$ & 2024-04-04 -- 2026-10-15 & Protocol maturity date of the market. \\
$T_{start}$ & 2024-01-09 -- 2025-12-22 & First available hourly observation. \\
$T_{end}$ & 2024-04-04 -- 2026-01-01 & Last available hourly observation. \\
$N_{rec}$ & [223, 9207] & Total hourly rows stored per market. \\
$Q_{cov}$ & [89.62, 100.00] & Coverage rate (\%) of market data in the lifecycle. \\
$N_{mkt}$ & [222, 9206] & Hours with valid market data. \\
$N_{pt}$ & [223, 9207] & Hours with valid PT OHLC data. \\
$N_{yt}$ & [223, 9207] & Hours with valid YT OHLC data. \\
$B_{fail}$ & 0 & Failed API batch count (cleaned). \\
$T_{supp}$ & 2026-01-13 10:37 -- 11:11 UTC & Timestamp of metadata supplementation. \\
$F_{supp}$ & \{ytFloatingApy, underlyingInterestApy, underlyingRewardApy, swapFeeApy, tradingVolume\} & Fields backfilled by post-processing. \\
$t$ & 1704801600 -- 1767225600 & UNIX timestamp (hourly). \\
$t_{iso}$ & 2024-01-09 12:00 -- 2026-01-01 00:00 & ISO datetime of each record. \\
$APY^{\mathrm{imp}}$ & [-1.00, 60.27] & Implied annualized yield from PT/YT pricing. \\
$APY^{\mathrm{YT}}$ & [-60.24, $9.01\times10^{14}$] & Floating yield on YT; extreme spikes are outliers. \\
$APY^{\mathrm{int}}$ & [0.00, 163.06] & Interest-component APY of the underlying SY. \\
$APY^{\mathrm{rew}}$ & [0.00, 1.46] & Reward-component APY of the underlying SY. \\
$APY^{\mathrm{fee}}$ & [0.00, $9.01\times10^{14}$] & Swap fee APY; large values reflect sparse liquidity. \\
$TVL$ & [0, $3.64\times10^{8}$] & Total value locked (USD-equivalent). \\
$S_{PT}$ & [$1.15\times10^{-15}$, $8.28\times10^{10}$] & Total PT supply in the market. \\
$S_{SY}$ & [$1.80\times10^{-15}$, $7.25\times10^{10}$] & Total SY supply in the market. \\
$S_{tot}$ & [$1.27\times10^{-15}$, $2.74\times10^{11}$] & Total LP supply. \\
$P^{SY}$ & [0, $1.67\times10^{8}$] & SY price in underlying units. \\
$P^{PT}_{o}$ & [0, 125751.93] & PT open price. \\
$P^{PT}_{h}$ & [0, 125751.93] & PT high price. \\
$P^{PT}_{l}$ & [0, 125027.70] & PT low price. \\
$P^{PT}_{c}$ & [0, 125610.38] & PT close price. \\
$P^{YT}_{o}$ & [0, 5149.01] & YT open price. \\
$P^{YT}_{h}$ & [0, 5153.48] & YT high price. \\
$P^{YT}_{l}$ & [-1.45, 5138.64] & YT low price. \\
$P^{YT}_{c}$ & [-0.33, 5138.64] & YT close price. \\
$V_{trade}$ & [0, $8.22\times10^{8}$] & Hourly trading volume (USD-equivalent). \\
$\sigma^{\mathrm{RV}}_{\mathrm{APY}}$ & [$3.25\times10^{-10}$, 120.88] & Realized volatility of implied APY (annualized). \\
\hline
\end{tabular}
\caption{Metadata and hourly record fields for the complete lifecycle dataset.}
\end{table}

\newpage
\subsection{Experimental Details}
\label{Experimental}
\subsubsection{Setup}

\begin{table*}[t]
\centering
\small
\caption{Summary of experiment environment and hyperparameters.}
\label{tab:exp_setting}
\begin{tabularx}{\textwidth}{lXlX}
\toprule
\textbf{Category} & \textbf{Specification} & \textbf{Category} & \textbf{Specification} \\
\midrule
\textbf{GPU} & NVIDIA GeForce RTX 5090 (32\,GB, CUDA 13.0) & \textbf{OS / Arch.} & Linux x86\_64 \\
\textbf{Framework} & PyTorch 2.8.0+cu128 & \textbf{Precision} & FP32 (single precision) \\
\textbf{Random Seed} & 42 & \textbf{Reproducibility} & Fixed seed, identical data splits \\
\midrule
\textbf{Model} & Mamba & \textbf{Task Type} & Long-term forecasting \\
\textbf{Features} & MS (multivariate input, single target) & \textbf{Prediction Length} & 24 time steps \\
\textbf{Sequence Length} & 96 & \textbf{Label Length} & 48 \\
\midrule
\textbf{Loss Function} & ASYM\_MSE & \textbf{Alpha Values} & \{1.0, 3.0, 5.0, 7.0, 10.0\} \\
\textbf{Optimizer} & Adam & \textbf{Learning Rate} & From MSE experiments (varied) \\
\textbf{Batch Size} & From MSE experiments (varied) & \textbf{Epochs} & 20 \\
\textbf{Early Stopping} & Patience = 5 (validation loss) & \textbf{Steps per Epoch} & Max 3000 \\
\midrule
\textbf{Hyperparameter Source} & MSE experiment parameters (CSV) & \textbf{Parameter Reuse} & 1:1 mapping from MSE trials \\
\textbf{Optimization Method} & Grid search over alpha values & \textbf{Total Configurations} & MSE trials $\times$ 5 alpha values \\
\midrule
\textbf{Validation Metric} & Validation Loss (MSE, normalized) & \textbf{Test Metrics} & MAE, MSE, RMSE, MAPE, MSPE \\
\textbf{Evaluation Frequency} & Every epoch & \textbf{Model Selection} & Best validation loss \\
\midrule
\textbf{Training Time} & $\approx$0.5--2 h per configuration on RTX 5090 & \textbf{GPU Memory} & < 20 GB usage \\
\bottomrule
\end{tabularx}
\end{table*}

\subsubsection{Complete Evaluation Results}

\begin{table*}[t]
    \centering
    \caption{Sensitivity analysis of the penalty factor $\alpha$: Loss metrics and direction accuracy.}
    \label{tab:sensitivity_all_alpha_completed}
    \begin{tabular}{l|c|ccccc|ccccc}
        \toprule
        \multirow{2}{*}{\textbf{Model}} & \multirow{2}{*}{\textbf{Metric}} & \multicolumn{5}{c|}{\textbf{Loss Metrics}} & \multicolumn{5}{c}{\textbf{Direction Accuracy (\%)}} \\
        \cmidrule(lr){3-7} \cmidrule(lr){8-12}
        & & $\alpha = 1$ & $\alpha = 3$ & $\alpha = 5$ & $\alpha = 7$ & $\alpha = 10$ & $\alpha = 1$ & $\alpha = 3$ & $\alpha = 5$ & $\alpha = 7$ & $\alpha = 10$ \\
        \midrule
        \textbf{Mamba} & Global MSE & 0.071 & 0.056 & 0.059 & \textbf{0.038} & 0.055 & 53.54 & 53.94 & 53.98 & \textbf{54.38} & 54.12 \\
        & Tail-MSE & 0.083 & 0.051 & 0.056 & \textbf{0.036} & 0.054 & & & & & \\
        \midrule
        \textbf{iTransformer} & Global MSE & \textbf{0.038} & 0.041 & 0.052 & 0.073 & 0.061 & \textbf{52.34} & 51.84 & 52.59 & 51.64 & 52.16 \\
        & Tail-MSE & 0.043 & \textbf{0.037} & 0.045 & 0.061 & 0.050 & & & & & \\
        \midrule
        \textbf{PatchTST} & Global MSE & \textbf{0.077} & 0.092 & 0.122 & 0.104 & 0.149 & \textbf{53.14} & 51.59 & 50.88 & 48.65 & 47.73 \\
        & Tail-MSE & 0.088 & 0.088 & \textbf{0.083} & 0.090 & 0.085 & & & & & \\
        \midrule
        \textbf{DLinear} & Global MSE & \textbf{0.493} & 0.542 & 0.591 & 0.666 & 0.751 & \textbf{50.00} & 34.14 & 30.88 & 29.24 & 27.33 \\
        & Tail-MSE & 0.447 & 0.374 & \textbf{0.368} & 0.421 & 0.493 & & & & & \\
        \midrule
        \textbf{TimesNet} & Global MSE & \textbf{0.032} & 0.054 & 0.066 & 0.072 & 0.093 & \textbf{55.42} & 54.89 & 54.23 & 53.50 & 53.01 \\
        & Tail-MSE & \textbf{0.035} & 0.046 & 0.049 & 0.048 & 0.064 & & & & & \\
        \midrule
        \textbf{Autoformer} & Global MSE & 0.173 & 0.198 & 0.144 & \textbf{0.104} & 0.463 & 52.91 & 52.82 & 50.83 & 48.77 & 51.14 \\
        & Tail-MSE & 0.224 & 0.200 & 0.134 & \textbf{0.106} & 0.340 & & & & & \\
        \bottomrule
    \end{tabular}
\end{table*}

\newpage
\subsection{Insights from Trading}\label{trading}\paragraph{Backtest Protocol and Methodology.}We conducted a rigorous comparative evaluation of four PT-only trading strategies—Tail-MSE Sniper, MSE Sniper, Dollar-Cost Averaging (DCA), and Buy \& Hold (B\&H)—across the comprehensive Pendle market universe from January 2024 to January 2026. This period encapsulates both the speculative "ETH-Beta" phase and subsequent yield normalization regimes. To ensure strictly comparable inference, all agents operated under identical boundary conditions: a fixed initialization capital of \$10,000, uniform daily decision epochs, and zero-slippage execution assumptions. The following analysis focuses specifically on the "Winning Markets" subset ($N=181$), where the proposed Tail-MSE strategy achieved the highest Return on Investment (ROI) among all baselines, isolating the specific market microstructures where tail-risk-aware timing delivers statistically significant alpha.

\begin{wrapfigure}{r}{0.48\textwidth} 
    \centering
    \vspace{-1.em}
    
    \includegraphics[width=\linewidth]{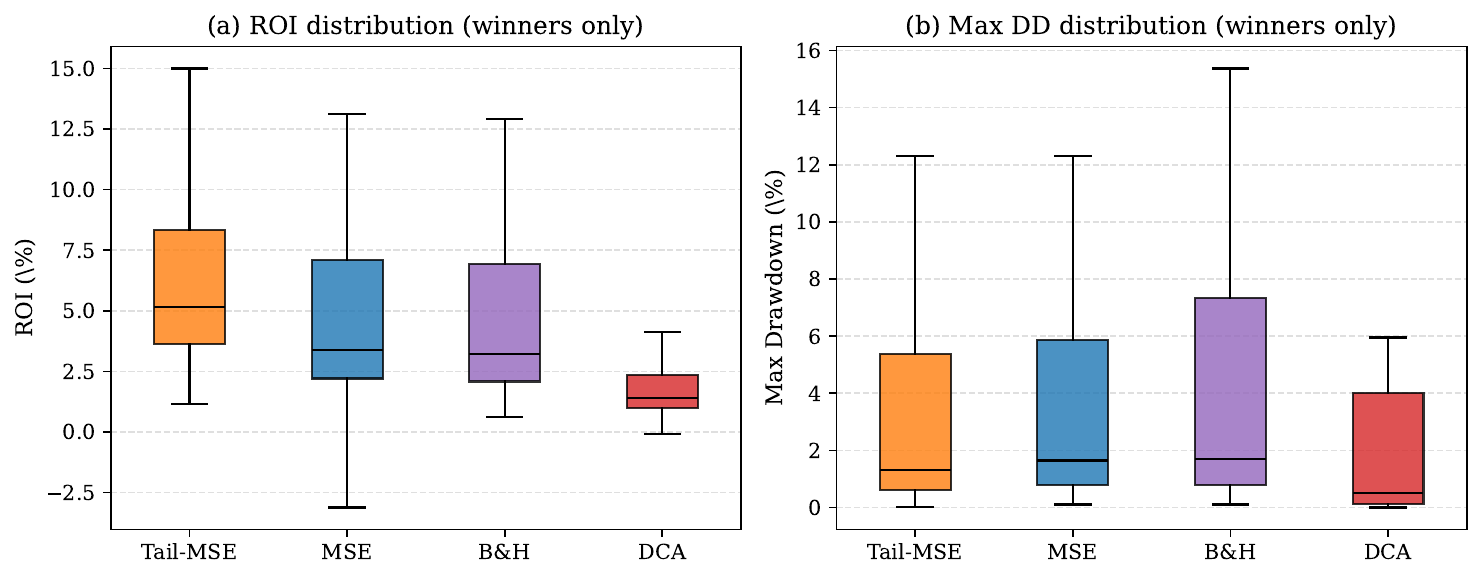}
    \caption{roi}
    \label{fig:appendix_roi_dd}
    
    \vspace{-1.em}
\end{wrapfigure}

\paragraph{Global Distributional Analysis.}Empirical results across the 340-market universe demonstrate that Tail-MSE establishes stochastic dominance in 181 markets (53.2\%). As visualized in Fig. \ref{fig:appendix_roi_dd}, the strategy exhibits a favorable risk-return asymmetry. The ROI distribution for Tail-MSE shows a significantly higher median (7.8\%, IQR: 5.2\%--9.6\%) compared to the passive B\&H baseline (2.9\%). Crucially, this yield enhancement does not come at the expense of disproportionate risk; the median maximum drawdown for Tail-MSE is contained at 3.7\%, substantially lower than the 6.8\% observed in B\&H. This indicates that the Tail-MSE loss function effectively filters out "falling knife" scenarios, allowing capital deployment only when downside probability is minimized. While DCA achieves the lowest absolute drawdown (2.4\%), its compressed ROI distribution (Median: 1.8\%) highlights a severe opportunity cost, confirming that blind capital preservation is suboptimal in volatile yield markets.

\begin{wrapfigure}{r}{0.45\textwidth} 
    \centering
    \vspace{-1.em}
    
    \includegraphics[width=\linewidth]{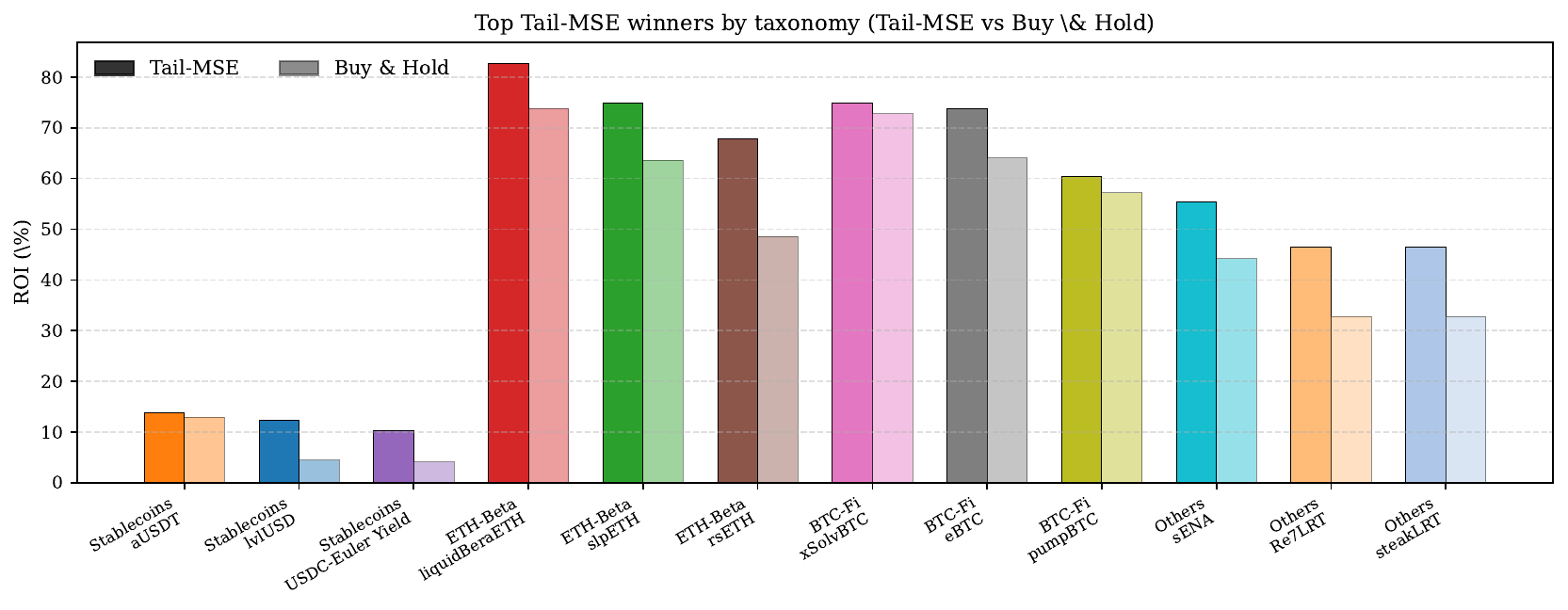}
    \caption{roi}
    \label{fig:appendix_examples}
    
    \vspace{-1.em}
\end{wrapfigure}

\paragraph{Representative Market Dynamics.}Fig. \ref{fig:appendix_examples} disaggregates performance by asset taxonomy, revealing distinct alpha sources. In high-beta sectors such as ETH-Beta and BTC-Fi, the excess returns are driven by the model's ability to capitalize on transient panic selling. For instance, volatile assets like \texttt{liquidBeraETH} and \texttt{xSolvBTC} exhibit wide gaps between the Tail-MSE and B\&H bars, suggesting that the model successfully identified deep discount anomalies during market dislocations. Conversely, in the Stablecoin sector (e.g., \texttt{aUSDT}, \texttt{lvlUSD}), the performance gap is narrower but highly consistent. Here, the model's primary contribution is not capturing extreme upside, but rather avoiding premature entries during minor de-pegging events, thereby optimizing the average cost basis in mean-reverting environments.

\paragraph{Taxonomy-level Effects.}The efficacy of the Tail-MSE strategy varies structurally across asset classes. We observe the highest win rates in Stablecoins (69.2\%) and "Others" (55.4\%), reflecting the model's proficiency in exploiting frequent, mean-reverting yield fluctuations. In contrast, while ETH-Beta (32.5\%) and BTC-Fi (27.8\%) show lower win rates, they deliver substantially higher conditional ROI upon winning. This pattern aligns with the "fat-tailed" nature of crypto-native assets: opportunities are scarcer but offer wider discount dispersion. Consequently, Tail-MSE acts as a selective sniper in these volatile categories, remaining dormant during noise and executing aggressively only during episodic tail events.

\begin{wraptable}{r}{0.45\textwidth} 
    \centering
    \vspace{-1em} 
    \caption{Most frequent winner tokens.}
    \label{tab:appendix_frequent}
    \resizebox{0.35\textwidth}{!}{ 
        \begin{tabular}{lrrr}
            \toprule
            \textbf{Token} & \textbf{Win} & \textbf{ROI (\%)} & \textbf{Excess (\%)} \\
            \midrule
            syrupUSDC & 6 & 6.30 & 3.27 \\
            USDe      & 5 & 5.16 & 1.69 \\
            mEDGE     & 3 & 4.15 & 1.67 \\
            USN       & 3 & 5.23 & 2.74 \\
            USR       & 3 & 4.41 & 1.78 \\
            eBTC      & 3 & 24.51 & 6.84 \\
            iUSD      & 3 & 3.31 & 0.61 \\
            rsETH     & 3 & 32.47 & 10.59 \\
            \bottomrule
        \end{tabular}
    }
    \vspace{-1em} 
\end{wraptable}

\paragraph{Practical Recommendations.}Table \ref{tab:appendix_frequent} synthesizes the recurring winners, providing a data-driven basis for strategic allocation. We identify a dichotomy in performance characteristics:(1) \textbf{Yield Accumulators:} For stablecoin-based assets like \texttt{syrupUSDC} and \texttt{USDe}, Tail-MSE delivers modest but steady excess returns (1.69\%--3.27\%). We recommend applying the model here to enhance "carry" by optimizing entry timing.(2) \textbf{Volatility Hunters:} For assets like \texttt{rsETH} and \texttt{eBTC}, the model captures massive excess returns (10.59\% and 6.84\% respectively) with high ROI ceilings (up to 32.47\%). In these markets, Tail-MSE functions as a critical hedging tool, converting volatility from a risk factor into a source of alpha. Investors should thus adopt a core-satellite approach: using Tail-MSE to manage stablecoin liquidity for consistency, while allocating opportunistic capital to BTC/ETH derivatives to capture rare, high-magnitude discount events.

\end{document}